\documentclass{article}
\usepackage{graphicx}
\usepackage{float}
\usepackage[a4paper, total={7in, 9in}]{geometry}
\usepackage{booktabs}
\usepackage[english]{babel}
\usepackage[square,numbers]{natbib}
\begin{document}

\title{Expectations vs Reality - A Secondary Study on AI Adoption in Software Testing}

\author{Katja Karhu, Jussi Kasurinen, Kari Smolander} 

\maketitle


\abstract[Abstract]{ In the software industry, artificial intelligence (AI) has been utilized more and more in software development activities. In some activities, such as coding, AI has already been an everyday tool, but in software testing activities AI it has not yet made a significant breakthrough. In this paper, the objective was to identify what kind of empirical research with industry context has been conducted on AI in software testing, as well as how AI has been adopted in software testing practice. To achieve this, we performed a systematic mapping study of recent (2020 and later) studies on AI adoption in software testing in the industry, and applied thematic analysis to identify common themes and categories, such as the real-world use cases and benefits, in the found papers. The observations suggest that AI is not yet heavily utilized in software testing, and still relatively few studies on AI adoption in software testing have been conducted in the industry context to solve real-world problems. Earlier studies indicated there was a noticeable gap between the actual use cases and actual benefits versus the expectations, which we analyzed further. While there were numerous potential use cases for AI in software testing, such as test case generation, code analysis, and intelligent test automation, the reported actual implementations and observed benefits were limited. In addition, the systematic mapping study revealed a potential problem with false positive search results in online databases when using the search string "artificial intelligence".

}

\renewcommand\thefootnote{}
\footnotetext{\textbf{Abbreviations:} AI, artificial intelligence; SQA, software quality assurance; ST, software testing.}

\renewcommand\thefootnote{\fnsymbol{footnote}}
\setcounter{footnote}{1}

\section{Introduction}\label{sec1}

In the past few years, modern artificial intelligence (AI) has brought changes to the field of software engineering. For example, code generation tools, such as GitHub Copilot, are already quite commonly used by software developers \cite{Mohamed2024ChattingChatGPT}. Some studies also indicate that code generation tools have boosted developer performance, and that the efficiency of software developers increases significantly with AI-based tools and platforms \cite{Cui2024TheCopilot, Goncalves2025AssessmentStudy}. In the software testing side, the interest in AI is high and growing, but AI adoption is lagging behind. In Perforce's \cite{Perforce2024TheTesting} industry survey, 48 percent of respondents indicated they were interested in AI but have not yet started any initiatives, and only 11 percent were already implementing AI techniques in software testing. Interest in AI is still growing a year later, in Perforce's newest 2025 industry survey \cite{Perforce2025TheTesting}, over 75 percent of survey respondents identified AI-driven testing as a pivotal component of their strategy for 2025. But actual adoption is still behind, with only 16 percent of respondents reported adopted AI in testing \cite{Perforce2025TheTesting}. 

According to the tertiary study by Amalfitano et al \cite{Amalfitano2023ArtificialStudy}, research on AI in software testing seems to have been conducted quite extensively already. This study \cite{Amalfitano2023ArtificialStudy}, conducted in 2023, is based on research published between 1995 and 2021. Given the recent advancements in generative AI and large language models (LLMs), we think it is very likely that research in this field has increased even more after 2021. 

Even though the research of AI in the software testing context has been extensive, Nguyen at al \cite{Nguyen-Duc2023GenerativeAgenda} found in 2023, that most of the existing studies on AI in software testing are "experimental studies and thus do not take into consideration the industrial context". King et al \cite{King2019AIPerspectives} had similar findings in 2019: "only a few of these works are backed by real-world case studies, or result in industrial tools and methods". 

Overall, there are interesting discrepancies related to AI adoption in software testing. Firstly, while there is extensive research on the application of AI to software testing, studies with an industry focus have been relatively rare. Secondly, although there is significant interest in the industry to utilize AI in software testing, the actual utilization rate appears to be quite low.

In this study, we want to investigate if the findings on the lack of studies with industrial context are still true. We are interested in finding what industry-related empirical research has been done recently on AI in software testing. This knowledge will help us find potentially new research directions, and hopefully help bridge the gap between industry and research. Our inclusion criteria is: original studies about AI adoption in the industry, with clearly described methodology, where data has been collected from via interviews, surveys (questionnaires) or other means from companies or experts, in the context of software testing.

Our research questions (RQ) are:
\begin{itemize}
    \item RQ1: What kind of studies have been made in the industrial or business context regarding AI adoption in software testing?
    \item RQ2: How is AI utilized in software testing in the industry?
\end{itemize}

With the first research question, we want to identify the current state of research into AI adoption in software testing. To find the answer to this question, we chose systematic mapping study as the method, because systematic mapping studies are well suited for giving an overview of a research area \cite{Petersen2015GuidelinesUpdate}.

The papers found by systematic mapping study will be then further analyzed using thematic analysis to answer the second research question, and to inductively find relevant themes related to the current state and issues of AI adoption in software testing. The goal of the second question is to increase the understanding about the current state of AI adoption in software testing. 

\section{Methods}\label{sec2}

As stated previously, the research questions dictated out choice of research methods. We utilized systematic mapping study, as described by Petersen et al \cite{Petersen2008SystematicEngineering, Petersen2015GuidelinesUpdate} to find answers to RQ1. Then, we performed a further detailed analysis of the papers using thematic analysis \cite{Braun2006UsingPsychology} to find answes to RQ2. 

\subsection{Systematic Mapping Study and Data Collection}

We followed the systematic mapping study process defined by Petersen et al \cite{Petersen2008SystematicEngineering}, which  involved defining the research question, conducting the search, screening of the papers, keywording (using abstracts), and data extraction and mapping process. According to Petersen et al \cite{Petersen2008SystematicEngineering}, the primary goal of a systematic mapping study is to provide an overview of a research area. This involves classifying and quantifying the type of research that has been done within the research area  \cite{Petersen2008SystematicEngineering}. Additional goals can be mapping the frequencies of the publications over time, or to identify forums where the research has been published  \cite{Petersen2008SystematicEngineering}. The research questions in systematic mapping studies are quite wide as they aim to discover research trends, such as what topics have been covered in existing literature \cite{Petersen2015GuidelinesUpdate}. 

\begin{table*}[!ht]%
\centering %
\caption{Phase 1: Failed initial search strings with number of results since 2020. \label{table_initial_search}}
\footnotesize\begin{tabular*}{\linewidth}{@{\extracolsep\fill}p{0.3\linewidth}p{0.27\linewidth}p{0.05\linewidth}p{0.05\linewidth}p{0.05\linewidth}}
\toprule
\textbf{Query {$^\dagger$}} & \textbf{Description} & \textbf{Google Scholar} & \textbf{Scopus} & \textbf{Semantic Scholar}\\
\midrule
("artificial intelligence" OR "generative ai" OR "large language model" OR "large language models") AND ("software testing" OR "quality assurance" OR "test automation") & Too many false positives. & 17600 & 30330 & 7\\\hline 
("artificial intelligence" OR "generative ai" OR "large language model" OR "large language models") AND\newline ("software testing" OR "quality assurance" OR "test automation") AND\newline ("case study" OR "survey" OR "empirical" OR "qualitative" OR " quantitative") AND\newline ("software engineering" OR "software development") & Added keywords to indicate that we are interested in the context of software development, and specifically in empirical studies & 17100  & 13015 & 251\\\hline 
("artificial intelligence" OR "generative ai" OR "large language model" OR "large language models") AND\newline ("software testing" OR "quality assurance" OR "test automation") AND\newline ("case study" OR "survey" OR "empirical" OR "qualitative" OR " quantitative") AND\newline ("software engineering" OR "software development") AND\newline ("sampling" OR "sample" ) & To further narrow down the studies, we want studies where sampling has been used, aka a specific sample has been studied. & 13000 & 2654 & 250\\\hline 
("artificial intelligence" OR "generative ai" OR "large language model" OR "large language models") AND\newline ("software testing" OR "quality assurance" OR "test automation") AND\newline ("case study" OR "survey" OR "empirical" OR "qualitative" OR " quantitative") AND\newline ("software engineering" OR "software development") AND\newline ("sampling" OR "sample" ) AND\newline (questionnaire OR interview) & Studies where data has been collected via questionnaires or interviews & 3960 & 198 {$^\ddagger$} & 250\\ 
\bottomrule
\end{tabular*}
\item[$^\dagger$] In Google Scholar format, with Scopus the query was adjusted for the Scopus format.
\item[$^\ddagger$] Dataset was selected for this study.
\end{table*}

One of our research goals in this study was to find articles about empirical research done on  AI adoption in software testing in the industry. Specifically, we were interested especially in empirical studies where people from the industry were interviewed or surveyed about AI adoption in software testing. The research question we wanted to explore was "What kind of studies have been made in the industrial or business context regarding AI adoption in software testing?" (RQ1). When comparing the methodology of systematic mapping studies and systematic reviews, systematic mapping study corresponded better with our research goal and question. We wanted to get an overview of what kind of research had been done previously, just what the systematic mapping study approach was designed to do. 

With research questions in place, we conducted the next step in the systematic mapping study process, the literature search. The databases we selected were Scopus, Google Scholar, and Semantic Scholar. We chose Semantic Scholar to see what the new AI-based research tools have to offer. We collected and filtered the data during October and November 2024. Google Scholar was especially important, because it contains also theses and and other grey literature which also features items of interest for us, works documenting industrial trials and implementation projects for new technology. According to Paez \cite{Paez2017GrayReviews} grey literature can be a rich source of evidence and it is important to be included systematic reviews, because of its potential to provide a balanced view of the evidence \cite{Paez2017GrayReviews}. We also performed a preliminary search for additional grey literature using DuckDuckGo. The results were mainly blog posts, but we noticed soon, that they did not fit our purposes: they did not document first-hand experiences, but rather summarized information, or they were focused on specific testing tools. Also, because of our choice of systematic mapping study, not systematic review, the requirement of finding all relevant studies was less stringent \cite{Kitchenham2010TheStudy}.

\begin{table*}[!ht]%
\centering %
\caption{Phase 2, new set of search strings \label{table_phase2_search_strings}}
\footnotesize\begin{tabular*}{\linewidth}{@{\extracolsep\fill}p{0.05\linewidth}p{0.3\linewidth}p{0.1\linewidth}p{0.2\linewidth}p{0.05\linewidth}p{0.05\linewidth}}
\toprule
\textbf{Dataset} & \textbf{Exact Query in DB Specific Format} & \textbf{DB} & \textbf{Description} & \textbf{Total} & \textbf{Included by Title}\\ 
\midrule
DS1 & 
( ALL (~"artificial intelligence"~OR "generative ai"~OR~"large language model"~OR~"large language models"~) AND ALL (~"software testing"~OR~"quality assurance"~OR~"test automation"~) AND ALL (~"case study"~OR~"survey"~OR~"empirical"~OR~"qualitative"~OR~"quantitative"~) AND ALL (~"software engineering"~OR~"software development"~) AND ALL (~"sampling"~OR~"sample"~) AND ALL (~questionnaire~OR~interview~) ) AND PUBYEAR {\textgreater}~2019~AND PUBYEAR {\textless}~2026 & Scopus & See table 1. & 198 & 87\\\hline 
DS2 & 
allintitle: "artificial intelligence"{\textbar}"generative ai"{\textbar}"large language model"{\textbar}"large language models" "software testing "{\textbar}"quality assurance"{\textbar}"test automation" & Google Scholar & Title contains AI and testing keywords.\newline Time limitation: articles since 2020 & 115 & 69\\\hline 
DS3 & 
( TITLE-ABS-KEY (~"artificial intelligence"~OR~"generative ai"~OR~"large language model"~OR~"large language models"~) AND TITLE-ABS-KEY (~"software testing"~OR~"quality assurance"~OR~"test automation"~) AND TITLE-ABS-KEY (~"case study"~OR~"survey"~OR~"empirical"~OR~"qualitative"~OR~"quantitative"~) AND ALL (~"software engineering"~OR~"software development"~) ) AND PUBYEAR {\textgreater}~2019~AND PUBYEAR {\textless}~2026 & Scopus & Title, abstract, and keywords contain AI, testing, and research methodology keywords.\newline Time limitation: articles since 2020 & 115 & 82\\\hline 

DS4 & 
"artificial intelligence in software testing"{\textbar}"AI in software testing" & Google Scholar & Full text search.\newline Time limitation: articles since 2020 & 200 & 184\\\hline 

DS5 & 
"artificial intelligence in software quality assurance"{\textbar}"AI in software quality assurance" & Google Scholar & Full text search.\newline Time limitation: articles since 2020 & 13 & 12\\\hline 

DS6 & 
"artificial intelligence in quality assurance"{\textbar}"AI in quality assurance" & 
Google Scholar & Full text search.\newline Time limitation: articles since 2020 & 48 & 7\\\hline 

DS7 & 
ALL (~"artificial intelligence in software testing"~OR~"AI in software testing"~OR~"artificial intelligence in software quality assurance"~OR~"AI in software quality assurance"~OR~"artificial intelligence in quality assurance"~OR~"AI in quality assurance"~) AND PUBYEAR {\textgreater}~2019~AND PUBYEAR {\textless}~2025 & Scopus & Full text search.\newline Time limitation: articles since 2020 & 42 & 32\\\hline 
 &  &  &  &  & 473\\\hline 
 & \multicolumn{4}{p{\dimexpr 0.811\linewidth-2\tabcolsep-\arrayrulewidth}}{
\textbf{Total after duplicate removal}} & 378\\ 
\bottomrule
\end{tabular*}
\end{table*}

We wanted to focus on the recent advancements by modern AI tools, such as Large Language Models and Generative AI, so the period is limited to articled published after 2020. Our search string development started with the basic terms: \textit{("artificial intelligence" OR "generative ai" OR "large language model" OR "large language models") AND ("software testing" OR "quality assurance" OR "test automation")}. But we soon ran to problems with false positive results when doing a full text search, as the papers found via the search in table \ref{table_initial_search} included quite many non-AI articles. In other words, articles that did not contain any of the AI search terms ("artificial intelligence", "generative ai", "large language model" or "large language models"). We tested the search string in three different databases, but in all three cases, the results seemed unusual. If we omitted the string "artificial intelligence" from the search, the number of results dropped by over 10 000 results from of the original in the case of Google Scholar. With Scopus the result was similar; search results number dropped significantly (for example, from 30 000+ to under 2500).

We found a few explanations for the significant number of non-AI papers. Artificial intelligence was mentioned 1) in the list of references, 2) in research institute/group names, 3) publication venue (e.g. journal, conference) name, or 4) unknown location, where the term "artificial intelligence" was not visibly mentioned at all within the paper. Cases one to three are normal and to be expected when doing a full text search, but the finding number four was something quite unexpected. This phenomenon would need further investigation. We suspect that in case 4, the generic AI statements included on the webpage where the papers are hosted might somehow cause false positive situations, or that the metadata in the papers contains the term artificial intelligence. It also could be that "artificial intelligence" is not the only search string that will bring false positive results. Finding the root cause or causes would require more investigation. 

These false positive results in literature searches may result in overestimating the amount of artificial intelligence research being done. In addition, they make the literature searches more cumbersome for researchers. 

In Semantic Scholar, the false positives presented themselves in a different way. With query 1, the resulting article amount was 7. However, when adding search terms, it seemed that Semantic Scholar started to show false positives, and the number of results was always ~250 articles after that. 

To get around the false positive results, we decided to use a different approach with the search strings. When using the term "artificial intelligence", we used two different strategies: 1) limited the search to title or title/abstract/keywords, and 2) combined artificial intelligence to another string to be able to do full text search (e.g. "artificial intelligence in software testing"). 

Only the 198 articles from Scopus were selected for the study from phase 1 search strings (see table \ref{table_initial_search}), since they seemed to contain valid results, and the dataset size was manageable. The Semantic Scholar results from table \ref{table_initial_search} did not seem reliable enough, so we decided to leave them out completely. Google Scholar, as well as the other Scopus results from table \ref{table_initial_search} were left out because of the amount of false positive articles.

In the screening of the papers, we had two primary exclusion rounds, round 1 (title) and round 2 (abstract). In addition, some papers were excluded in the keywording phase (that was also the starting point of the thematic analysis), when they were explored in even more detail. In the end the total number of relevant papers was 17. The language used in most of the studies was English, but we also had one study in Finnish  \cite{Ahven2022UtilizationAssurance}.

The exclusion round 1 was done based on paper titles according to our exclusion criteria (see table \ref{table_exclusions}). For example, literature reviews, papers without software engineering context, non-AI papers, papers about the testing of AI systems, etc were excluded. If the title of the paper was clearly out of our scope, we excluded it. Unclear cases went ahead to exclusion round two, where we evaluated the abstract (or in unclear cases, the whole paper). The column "Included by Title" in table \ref{table_phase2_search_strings}, we can see how many papers were included in round one of exclusions. In the exclusion round 2, we excluded the papers based on abstracts. If by after reading the abstract the applicability of the paper was inconclusive, the abstract, the rest of the paper was explored until a decision could be made. Usually going through the introduction and conclusions as suggested by Petersen et al \cite{Petersen2008SystematicEngineering} was enough to make this decision. 

\begin{table*}[!ht]%
\centering %
\caption{Exclusion Criteria \label{table_exclusions}}%
\footnotesize\begin{tabular*}{\linewidth}{@{\extracolsep\fill}p{0.30\linewidth}p{0.65\linewidth}@{\extracolsep\fill}}
\toprule
\textbf{Exclusion criteria} & \textbf{Reason}\\
\midrule
Literature studies & We were interested only in original studies \\\hline 
Wrong domain &  For example articles about quality assurance in medicine and the manufacturing industry were left out \\\hline 
No practical/industry context & Studies did not involve software development organizations or experts \\\hline 
No software testing context & For example, Use of AI for education and research, or job market related studies. We were interested in studies made in industry related to software testing and quality assurance of software. \\\hline 
Testing of AI systems & In this study we were interested in the adoption of AI in software testing, not the the testing of AI systems \\\hline 
Experimental studies with technology focus & E.g. comparison between different AI tools, or development of new technological solutions were excluded. \\\hline 
Developer focus & We chose to focus only on the software testing and QA specialists. \\\hline 
Data not collected from humans & E.g. data collected from code repositories \\\hline 
Language & Language was other than English, Finnish or Swedish \\ 
\bottomrule
\end{tabular*}
\end{table*}

In the process by Petersen et al \cite{Petersen2008SystematicEngineering} keywording is done by abstract. Since the number of papers was quite manageable, at around 20-30 at this point, we did the keywording based on the entire contents of the papers. Here, we used the tagging feature in Mendeley reference manager to manage our observations about the topics featured in the paper, research methodology, type of paper (grey literature, peer-reviewed), and geographical location. At this point, we also found when reading the full papers that some still needed to be excluded according to our exclusion criteria.

The final phase of the systematic mapping study process is the data extraction and the mapping of the studies. Because our sample size, the number of papers, was very small, a further statistical analysis was not seen as very useful, since the results would not be statistically significant. Therefore, instead of generating a systematic map as the result as is the custom \cite{Petersen2008SystematicEngineering}, the resulting comparison of the papers was done in a simpler table format. We classified the papers based on publication year, research and data collection methods, and whether the papers were grey literature or peer-reviewed. The results are described in more detail in the Results -section of the paper. 

\subsection{Thematic Analysis}
 
 In this study, we selected the qualitative approach with thematic analysis (TA) as the data analysis method, because we wanted to gain a deeper understanding of the topic: to find explanations on how AI is utilized in software testing (RQ2), to find additional differences and commonalities between the selected empirical studies, and to also find new potential unexplored (or less explored) research areas. Braun and Clarke \cite{Braun2006UsingPsychology, Braun2023TowardResearcher} define thematic analysis as a flexible qualitative analysis method, or more appropriately, a family of  methods, "for identifying, analyzing, and reporting patterns (themes) within data".  A theme is a concept that captures important patterned information and insights about the data, related to the research question \cite{Braun2006UsingPsychology}. Because of it's flexible nature, thematic analysis can be applied across a range of epistemological approaches \cite{Braun2006UsingPsychology}. This also means that researchers have also the responsibility of being transparent, and disclosing their underlying position and assumptions  \cite{Braun2019ReflectingAnalysis}

 Braun and Clarke \cite{Braun2023TowardResearcher} have further classified the thematic analysis family of methods into four different approaches: coding reliability, codebook and reflexive TA, and thematic coding. These approaches can be divided into small q and big Q approaches \cite{Braun2023TowardResearcher}. Small q approaches operate within the framework of (post)positivism \cite{Braun2023TowardResearcher}. An example of this is the coding reliability approach, that emphasizes procedures for ensuring the objectivity, reliability or accuracy of coding and  minimizing researcher bias. On the other side, reflexive thematic analysis is a Big Q approach \cite{Braun2023TowardResearcher}. It is non-positivist, and "embraces researcher subjectivity as a resource for research" \cite{Braun2023TowardResearcher}. In reflexive TA researcher subjectivity is embraced as a resource, and overall the practice is subjective: coding can never be accurate, as coding is an interpretive practice \cite{Braun2023TowardResearcher}. Codebook TA combines the procedures of coding reliability with some of the features of reflexive TA \cite{Braun2023TowardResearcher}. In thematic coding, the grounded theory coding procedures are used to develop themes from data \cite{Braun2023TowardResearcher}.
 
 The approach we selected for our study was reflexive TA, we later refer to it as "thematic analysis" in this study. Reflexive TA was selected because it is in line with the researchers' non-positivist views of the software engineering practice. Also, it is a lesser used thematic analysis approach in the area of empirical software engineering, hopefully pro. 

 Additionally, thematic analysis can be deductive or inductive: themes can be identified from the data inductively, or they can come from existing theories (deductive approach) \cite{Braun2006UsingPsychology}. In the inductive approach, the themes are strongly linked to the data, and the themes do not necessarily belong to any pre-existing coding frames \cite{Braun2006UsingPsychology}. Our approach was inductive and data-driven. We did not use any theoretical framework in generating our themes or codes: they emerged from the data. 
 
 According to Braun and Clarke \cite{Braun2023TowardResearcher} "positivism-creep" is the most common problem in thematic analysis. It is a situation where "positivism slips unknowingly into reflexive TA"  \cite{Braun2023TowardResearcher}. Examples of positivism-creep can be for example using concepts such as researcher bias, or emphasizing the reliability of coding \cite{Braun2023TowardResearcher}. Another potential pitfall in reflective TA is topic summaries as themes instead of capturing and interpreting a core idea or meaning \cite{Braun2023TowardResearcher}. If a theme has been developed before analyzing the data, or it maps closely to the research question, or it has a one-word name that identifies the topic, such as "Experiences of...", "Barriers to...", it is most likely a topic summary.
   
Braun and Clarke \cite{Braun2006UsingPsychology} state that another very important decision to make in thematic analysis is to specify, on which level the themes are identified. The analysis usually focuses on one of the levels: semantic or latent. In the semantic approach, the themes are explicit, aka the researchers does not try to interpret beyond the surface meaning of the data \cite{Braun2006UsingPsychology}. When the data has been organized, patterns have been identified, and summarized, it is then possible to progress into interpretation and to theorize on the meanings and implications based on the patterns \cite{Braun2006UsingPsychology}. In the latent approach, the analysis is interpretive and theorized right from the start \cite{Braun2006UsingPsychology}. The researcher's goal is to interpret underlying meanings, assumptions, concepts and ideologies that influence the semantic content of the data \cite{Braun2006UsingPsychology}. In this study, we applied mostly semantic (explicit) coding, because as a secondary study, we were missing a lot of the context associated with the data (e.g. original interview transcripts or survey answers), so latent coding was quite difficult to accomplish. 

We focused the analysis mostly on the result sections of the papers, because we wanted to analyze the first-hand empirical observations from the papers, especially for the "actual use cases" and "actual benefits". From the discussion and conclusions section we also picked up potential use cases and benefits. If a benefit or a use case was mentioned only in the previous literature (usually in introduction and background sections), we did not include it in our analysis
  
 In the analysis we followed the phases of thematic analysis by Braun and Clarke \cite{Braun2006UsingPsychology}:
\begin{itemize}
    \item Phase 1: familiarizing yourself with your data
    \item Phase 2: generating initial codes
    \item Phase 3: searching for themes
    \item Phase 4: reviewing themes
    \item Phase 5: defining and naming themes
    \item Phase 6: producing the report
\end{itemize}

It is worth noting, that the process is not as linear as it seems based on the list and the consecutive phases. Writing and analysis in qualitative research is inter-vowen, so in practice producing the report starts already in phase 1 \cite{Braun2012ThematicAnalysis}. In the beginning, informal notes and memos are an important tool documenting your process and ideas \cite{Braun2012ThematicAnalysis}.

The purpose of the first phase is to generate an initial list of points of interest in the data \cite{Braun2006UsingPsychology}. The first phase went hand-in-hand with our systematic mapping study. The first phase of thematic analysis shared commonalities with the keywording phase of the systematic mapping study, where we used the tagging feature in Mendeley reference manager to note down high level observations about the contents of the papers. This first phase already generated some high level codes that later evolved into categories (for example "benefits"). For the later phases of analysis, we used a tool called NVivo 14 to code the contents of the papers. In the second phase, we mainly utilized descriptive coding and in-vivo coding. 
 
The goal of the third phase is to combine different codes with shared meanings into themes or sub-themes \cite{Byrne2022AAnalysis}. We started to group the codes into potential themes, such as "risks", "use cases", "benefits". We went through all the codes we had created so far, and gave them a prefix that indicated their greater theme. During this phase, we noticed that some of the papers \cite{Purovesi2024TestAI, Layman2024GenerativeTesting, Barenkamp2020ApplicationsEngineering} raised the issue of having a gap between expectations and "reality": the expectations were very high, but actual implementations were lagging behind. For example, Barenkamp et al \cite{Barenkamp2020ApplicationsEngineering} stated that there is a  "divergence between the theoretical understanding of what AI could possibly achieve and the practical capabilities of available AI programs". Similarly, Purovesi \cite{Purovesi2024TestAI} had found that "there is an emerging conflict between expectations and reality, and a need for clarity and understanding of the potential and limitations of AI". 
 
Reviewing the potential themes (phase four) is a recursive process, where "the themes are reviewed in relation to the coded data and the entire data set" \cite{Braun2012ThematicAnalysis}. The key questions to ask at this point are \cite{Braun2012ThematicAnalysis}: 
\begin{itemize}
    \item Is this a theme (it could be just a code)?
    \item If it is a theme, what is the quality of this theme (does it tell me something useful about the data set and my research question)?
    \item What are the boundaries of this theme (what does it include and exclude)?
    \item Are there enough (meaningful) data to support this theme (is the theme thin or thick)?
    \item Are the data too diverse and wide ranging (does the theme lack coherence)?
\end{itemize}

When we reached phase four, it was clear that "expectations vs reality" became the central theme in our analysis. We realized that our dataset had the potential of providing more information about this phenomenon, and decided to investigate and analyze this difference further. We noticed that the "use cases" and "benefits" themes that had been identified earlier indeed contained these two different concepts of expectations and reality. Additionally, in order to answer our RQ2, we needed to find out how AI is actually utilized in software testing in practice in any case, which was aligned with the actual use cases documented in the data.The newly created categories in the under use cases and benefits then were:
 \begin{itemize}
     \item Actual use cases: how is AI used in reality in the software testing context (answers RQ2)
     \item Expected use cases: what kind of potential use cases there are for AI in the software testing context 
     \item Actual benefits: what kind of benefits from AI in software testing have been observe in reality
     \item Expected benefits: what kind of benefits are expected from AI in software testing
 \end{itemize}

In phase 5, defining and naming themes, you need to identify and clearly state, what is unique and specific about each theme \cite{Braun2012ThematicAnalysis}. Good themes are easy to sum up in a few sentences, do not overlap, directly address your research question, and have a clear focus, scope and purpose \cite{Braun2012ThematicAnalysis}. According to Braun and Clarke \cite{Braun2012ThematicAnalysis} you may also want to have sub-themes when there are overarching patterns within the data. In our case, "use cases" and "benefits" became the sub-themes of "expectations vs reality" in the end. 

\begin{figure*}
    \centerline{
    \includegraphics[width=\textwidth]{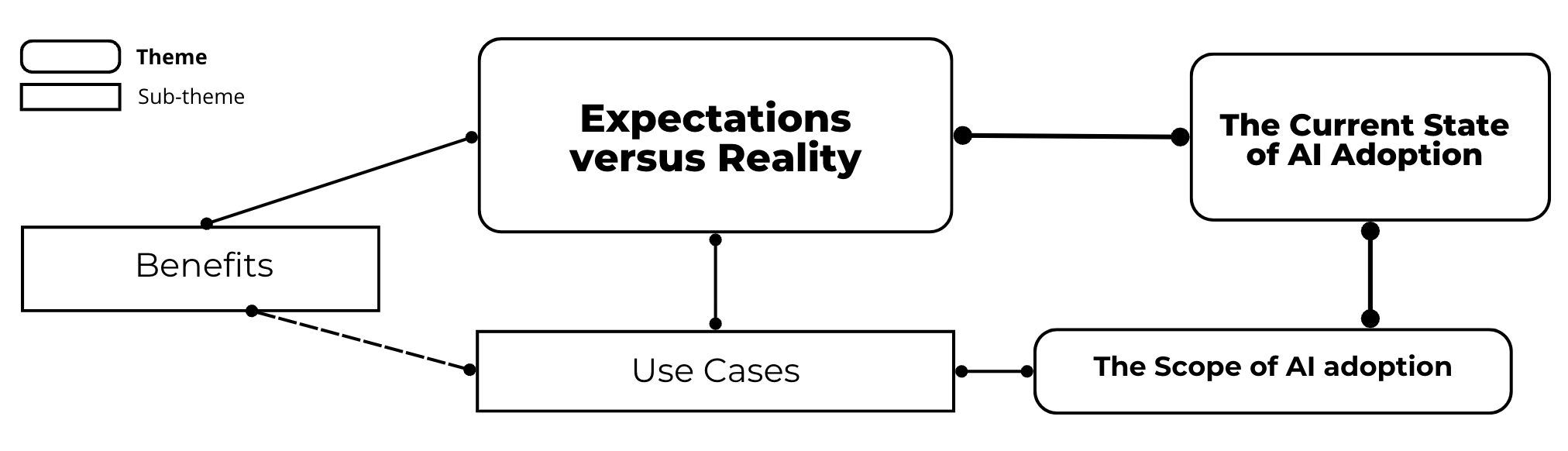}}
    \caption{The thematic map of identified themes and sub-themes}
    \label{fig:thematic-map}
\end{figure*}

The thematic map \ref{fig:thematic-map} displays the relationships between our themes and sub-themes. Expectations vs reality was our main theme, under it were the sub-themes use cases and benefits, which were further grouped in to actual and expected -categories. The "current state of AI adoption in software testing" was our supporting theme related, which compounded background information of current state of AI adoption from the studies we analyzed. The scope of AI adoption was another additional theme that we identified when doing a deeper analysis of the use cases.

Writing the report in phase 6, and the analysis in general, was more of an iterative process, where especially phases four to six were repeated whenever new ideas and viewpoints came up. For example, the theme "the scope of AI adoption" was formed only after a very detailed analysis of the use case sub-theme was done. According to Braun and Clarke \cite{Braun2012ThematicAnalysis}, the purpose of the report is to provide a compelling story about the data, based on your analysis. The goal is to go beyond pure description and to make an argument that answers the research question \cite{Braun2012ThematicAnalysis}. 

We also found other interesting major themes, such as risks of AI, barriers of AI adoption, and  attitudes towards AI. These themes went through the phases one to five in the thematic analysis process, but problems arose in the reporting at phase six. Trying to fit all the interesting themes to one paper quickly became quite daunting, the complexity and the length of the paper kept growing. The task of having to compare the all the results with additional literature providing a coherent Discussion section was looming in the distance. The reporting of the themes felt superficial and there was no common thread that would run through all the themes. 

Selecting the right number of themes for your study is a balancing act: too many themes make the analysis overly complex and incoherent, and too few themes can result in a failure to fully explore the data \cite{Byrne2022AAnalysis}. Thematic analysis itself does not specify how many themes, categories, or codes you should have and should report \cite{Byrne2022AAnalysis}. One of the root causes for too many themes to report was that we had too many research questions for one study. By cutting down the research questions to two, the number of themes to report was reduced to three. In addition, we went through the remaining themes by identifying the ones that best served our research question (RQ2). In the end, three themes remained: "expectations vs reality", "current state of AI adoption in software testing", and "the scope of AI adoption". Having less themes allowed us to deepen the analysis of the selected themes, which would have been quite superficial otherwise. 

\section{Results}\label{sec3}

We used systematic mapping study to find answers to our first research question: what kind of studies have been made in the industrial or business context regarding AI adoption in software testing? We then further analyzed the found articles using thematic analysis to answer the second research question: how is AI utilized in software testing in the industry?

\subsection{Systematic Mapping Study Results }

The short answer to our RQ1 (What kind of studies have been made in the industry regarding AI adoption in software testing?) can be seen in tables \ref{table_number_of_papers_per_type}, \ref{table_number_of_papers_per_year}, and \ref{table_paper_methodology}. Nine of studies were peer-reviewed, six were theses, and the remaining two were a report \cite{Hossain2024AICompanies} and a virtual round-table \cite{Layman2024GenerativeTesting}. Over half of the papers were published in 2024. However, from the studies published in 2024, only three were peer-reviewed. This was the same number of peer-reviewed studies as in 2023. It seems that studies with industry context are still quite rare. It is worth noting that the variety of research methodologies of the studies we selected (see Table \ref{table_paper_methodology}) was partially constrained by our research interest and the exclusion criteria. 
 
Qualitative research was the most used approach in the papers we found. In about half of the qualitative studies the used data analysis process or methodology was not described in detail. When the methodology was mentioned, it was usually thematic analysis. But it was also interesting to see a cyber ethnography study, as well as an action research study. We also saw multiple studies using mixed methodology (see table \ref{table_paper_methodology}), for example, by combining a statistical analysis of a survey with a qualitative analysis of open questions or additional interviews. 

Regarding the thematic analysis studies, two of them specified, which variant of thematic analysis was used, and in both cases it was the approach described by Cruzes and Dybå \cite{Cruzes2011RecommendedEngineering}. Cruzes and Dybå's guidelines for thematic analysis represent the more (post)positivist school of though, where the goal is to minimize researcher bias and maximize the validity of the study. In most cases the research philosophy behind the studies seemed to be positivist or, at least in one case, pragmatist \cite{Amarasekara2023ChallengesProcess}, although the position was not usually explicitly stated. Some of the qualitative studies did however explain the reasoning behind the qualitative approach. The main reasons could be summarized as increasing understanding of the studied phenomenon, or due to the social setting of the research.

Data collection was mainly performed via survey questionnaires or interviews (see table \ref{table_paper_data_collection}). One of our "outlier" papers was a virtual round-table with experts \cite{Layman2024GenerativeTesting}. It did not fall any of the formal research methodology categories, but it did still provide us with interesting first-hand information about the latest issues and trends in AI assisted testing. It's worth noting that some studies mentioned only the data collection method, but not data analysis method, or vice-versa. 

\begin{table*}[!htbp]%
\centering %
\caption{Number of papers per type \label{table_number_of_papers_per_type}}%
\footnotesize\begin{tabular*}{\linewidth}{@{\extracolsep\fill}lp{0.3\linewidth}@{\extracolsep\fill}}
\toprule
\textbf{Type} & \textbf{Count} \\
\midrule
Peer-reviewed (journal or conference) & 9 \cite{Amalfitano2024AIPractitioners, Barenkamp2020ApplicationsEngineering, Bhuvana2023IntegrationDevelopment, Gutierrez2020AI-PoweredDevelopment, Khan2024AI-BasedTesting, Kumar2023TheChallenges, Qazi2023SoftwarePakistan, Ramchand2022RoleAssurance, Santos2024AreTesting}\\\hline 
Thesis & 6 \cite{Adu2024ArtificialTechniques, Ahven2022UtilizationAssurance, Amarasekara2023ChallengesProcess, Jauhiainen2024ArtificialTesting, Laine2024ATESTING, Purovesi2024TestAI}\\\hline 
Other grey literature & 2 \cite{Hossain2024AICompanies, Layman2024GenerativeTesting}\\\hline 
\textbf{Total:} & 17\\ 
\bottomrule
\end{tabular*}
\end{table*}

\begin{table*}[!htbp]%
\centering %
\caption{Number of papers per year \label{table_number_of_papers_per_year}}%
\footnotesize\begin{tabular*}{\linewidth}{@{\extracolsep\fill}lp{0.3\linewidth}@{\extracolsep\fill}}
\toprule
\textbf{Year} & \textbf{Count} \\
\midrule
2020 & 2 \cite{Gutierrez2020AI-PoweredDevelopment, Barenkamp2020ApplicationsEngineering} \\\hline 
2021 & 0 \\\hline 
2022 & 2 \cite{Ramchand2022RoleAssurance, Ahven2022UtilizationAssurance} \\\hline 
2023 & 4 \cite{Bhuvana2023IntegrationDevelopment, Kumar2023TheChallenges, Qazi2023SoftwarePakistan, Amarasekara2023ChallengesProcess}\\\hline 
2024 & 10 \cite{Amalfitano2024AIPractitioners, Khan2024AI-BasedTesting, Santos2024AreTesting, Adu2024ArtificialTechniques, Jauhiainen2024ArtificialTesting, Laine2024ATESTING, Purovesi2024TestAI, Hossain2024AICompanies, Layman2024GenerativeTesting} \\\hline 
\textbf{Total:} & 17\\ 
\bottomrule
\end{tabular*}
\end{table*}

\begin{table*}[!htbp]%
\centering %
\caption{Research methodologies utilized in the papers \label{table_paper_methodology}}%
\footnotesize\begin{tabular*}{\linewidth}{@{\extracolsep\fill}lp{0.3\linewidth}@{\extracolsep\fill}}
\toprule
\textbf{Data Analysis Method} & \textbf{Studies} \\
\midrule
Quantitative &  6 \cite{Gutierrez2020AI-PoweredDevelopment, Santos2024AreTesting, Amarasekara2023ChallengesProcess, Ramchand2022RoleAssurance, Qazi2023SoftwarePakistan, Amalfitano2024AIPractitioners} \\\hline 
Qualitative &  13 \cite{Laine2024ATESTING, Khan2024AI-BasedTesting, Gutierrez2020AI-PoweredDevelopment, Barenkamp2020ApplicationsEngineering, Santos2024AreTesting, Adu2024ArtificialTechniques, Amarasekara2023ChallengesProcess, Bhuvana2023IntegrationDevelopment, Ramchand2022RoleAssurance, Qazi2023SoftwarePakistan, Purovesi2024TestAI, Kumar2023TheChallenges, Ahven2022UtilizationAssurance} \\\hline 
    \hspace{3mm}Thematic analysis &  4 \cite{Gutierrez2020AI-PoweredDevelopment, Santos2024AreTesting, Amarasekara2023ChallengesProcess, Ahven2022UtilizationAssurance} \\\hline 
    \hspace{3mm}Action research & 1 \cite{Adu2024ArtificialTechniques} \\\hline 
    \hspace{3mm}Cyber Ethnography & 1 \cite{Khan2024AI-BasedTesting} \\\hline 
    \hspace{3mm}Case Study & 4 \cite{Gutierrez2020AI-PoweredDevelopment, Purovesi2024TestAI, Ahven2022UtilizationAssurance, Laine2024ATESTING} \\\hline
Mixed method &  5 \cite{Amarasekara2023ChallengesProcess, Gutierrez2020AI-PoweredDevelopment, Ramchand2022RoleAssurance, Santos2024AreTesting, Barenkamp2020ApplicationsEngineering} \\
\bottomrule
\end{tabular*}
\end{table*}

\begin{table*}[!htbp]%
\centering %
\caption{Data collection methods used in the papers \label{table_paper_data_collection}}%
\footnotesize\begin{tabular*}{\linewidth}{@{\extracolsep\fill}lp{0.3\linewidth}@{\extracolsep\fill}}
\toprule
\textbf{Data Collection Method} & \textbf{Studies} \\
\midrule
Survey (questionnaire) & 8 \cite{Hossain2024AICompanies, Amalfitano2024AIPractitioners, Gutierrez2020AI-PoweredDevelopment, Santos2024AreTesting, Jauhiainen2024ArtificialTesting, Amarasekara2023ChallengesProcess, Ramchand2022RoleAssurance, Qazi2023SoftwarePakistan}  \\\hline 
Interview & 10 \cite{Laine2024ATESTING, Hossain2024AICompanies, Barenkamp2020ApplicationsEngineering, Adu2024ArtificialTechniques, Amarasekara2023ChallengesProcess, Bhuvana2023IntegrationDevelopment, Qazi2023SoftwarePakistan, Purovesi2024TestAI, Kumar2023TheChallenges, Ahven2022UtilizationAssurance} \\\hline
Data from online sources (cyber ethnography) & 1 \cite{Khan2024AI-BasedTesting} \\\hline
Virtual round-table & 1 \cite{Layman2024GenerativeTesting} \\
\bottomrule
\end{tabular*}
\end{table*}

\subsection{Thematic Analysis Results}

With thematic analysis, we found more answers than we initially looking for, as the inductive analysis revealed a plethora of interesting themes. We could not report all the themes in one paper, so we had to focus on one research question and the themes related to it. Our second research question (RQ2) that we wanted to explore using thematic analysis was: how is AI utilized in software testing in the industry? 

This research question lead us first to the use cases: what is done in the industry. Next, we also wanted to evaluate the benefits: how useful is AI in software testing. This was done by finding the reported benefits, and also connecting the benefits to specific use cases. We also found evidence of a difference between expectations and reality, as there was mismatch between expected and actual use cases, as well as the actual benefits and and expected benefits.

Based on our findings, there are at least three different answers to RQ2. First is the listing of the use cases from the industry. Secondly, we identified different scopes of AI adoption: individual-level, where individuals use AI as an assistant to complete their tasks, and system-wide, where AI is used to automate large system-wide tasks. Thirdly, "not ideally", as AI adoption in software testing still seems to be in very early stages, and the benefits drawn from the use cases can be vague.

\subsubsection{Expectations versus Reality: Use Cases}

The studies contained information about multiple actual and potential use cases for AI in software testing. They were described at varying levels of detail, as some were only mentioned, and some where described in a detailed manner. In some cases also the the benefits that resulted or were expected result from the use case, as well as challenges related to the use case, were described. 

 We grouped (table \ref{table_use_case_expectations_reality}) the use cases into categories "Generation", "Analysis",  "Core Testing Activities", "Prioritization", "Repair", and "Test Maintenance and Infrastructure".  Table \ref{table_use_case_expectations_reality} contains the list of both actual use cases and expected use cases grouped in these categories. As can be seen from the variety of use cases, testing contains much more than the "Core Testing Activities". 

\textbf{Generation}. In the category "Generation" we can find use cases related to generating code or text. Test case generation, along with code generation, was the most often mentioned actual use case for AI in software testing in the papers included in this study. We found that as a term, test case generation contains a lot variety. In general, test case generation can be done in many levels, from unit tests to integration and UI testing. Test cases can also be manual or automated. Also the starting point of test case generation can vary. The test cases were generated from code, bug reports, user stories, or other natural language descriptions. As concepts, the use cases "code generation" and "test case generation" can often overlap, since the "test case" can actually be the also code, for example in the unit testing and test automation contexts. Also, the systems used in test case generation can be custom solutions developed in-house, or commercial tools. The solutions may also be highly technical, such as specific algorithms. In the study by Ahven {\cite{Ahven2022UtilizationAssurance}, an interviewee described a proof-of-concept (POC) where bug reports were converted into automated test cases. Amalfitano et al \cite{Amalfitano2024AIPractitioners} found that in the context of GUI testing, respondents used commercial tools like "testRigor" to generate tests scripts via natural language, but also neural networks and deep learning algorithms, as well as NLP techniques to create user stories, which were translated by manual scripting to test cases. Also, there seemed to be two ways how test case generation was implemented: on individual-level, where QA specialists use tools to generate test cases in a smaller scope, or system-wide level, where the test cases were generated on a large scale.  It would be interesting to investigate the AI-driven test case generation in more detail in the practical setting, to more deeply understand how and in what context it test case are generated.   

Regarding expectations, Santos et al \cite{Santos2024AreTesting} found that test case generation was the most anticipated testing activity where LLMs could be applied, especially among users who do not utilize LLMs. In the study by Jauhiainen \cite{Jauhiainen2024ArtificialTesting}, 18 out of 25 respondents said that "test case generation is the issue that they wish AI could help with or solve totally". On the other hand, there was skepticism towards test generation by AI. In Adu's \cite{Adu2024ArtificialTechniques}, some of the respondents were very positive towards test case generation, but some were concerned whether the generated tests are useful, as in do they test the correct things, and does AI have enough domain knowledge to generate tests. These respondents viewed that human supervision in AI-based test case generation is needed \cite{Adu2024ArtificialTechniques}. 

Unsurprisingly, code generation was another common actual use case. Code generation was used especially in the test automation context. As a term, code generation was not as complex as "test case generation". It contained activities, such as boilerplate code generation, generating methods, test script generation. Code generation was often associated with code analysis as well. Unlike AI-based test case generation, which was also described as a system-wide activity, code generation with AI was a more individual activity, as these quotes from the studies indicate: 

\textit{"I have used an LLM to help generate code for some functions that I was stuck on."} \cite{Santos2024AreTesting}

\textit{"I have used some large language models. I have attempted to use them for analyzing or generating code."} \cite{Adu2024ArtificialTechniques}

\textit{"What I've tried and used myself, if I start to write a python script related to test automation, and if I know roughly what I would like to do, but then it would take a lot of time to plan it, I have noticed that, for example, even ChatGPT gives pretty good basis and ideas to that what I should do."} \cite{Purovesi2024TestAI}. 

Test data generation was after test case generation the second most popular issue AI could help with or solve totally in Jauhiainen's \cite{Jauhiainen2024ArtificialTesting} study. However, there were not many details available in any of the studies on how QA specialists currently utilize AI in test data generation. In Purovesi's \cite{Purovesi2024TestAI} study, it was mentioned that "generating test data with AI remains a challenge, as it can only generate basic data". On the other hand, in Khan et al's \cite{Khan2024AI-BasedTesting} study it was mentioned that "through the automated generation of test data that simulates a variety of scenarios and settings, AI-based software testing can assist in improving test data management". Without any the detailed knowledge of test data generation with AI in a practical setting, it is difficult to analyze the issues further. There seems to be room for empirical studies in the industry regarding test data generation and management.

Testing also involves having to document a multitude of things. The actual use cases in document generation were related to test plan generation, but did not provide much additional details. One participant in the study by Santos et al \cite{Santos2024AreTesting} indicated that \textit{"LLMs can provide standardized test plan templates or frameworks, ensuring that important sections like objectives, scope, resources, and timelines are included in the plan"}, but it was unsure if they were talking about actual or expected use cases.

 In the studies, participants expected AI to help in in various document generation activities, such as test plans, test strategy, user guidelines, user stories, standard and security related documentation, and other testing related documentation. Laine \cite{Laine2024ATESTING} found that AI-based tool, used in the company he investigated, "could aid test engineers with writing test plans and testing strategy documents and summarize information from long documents saving significant amounts of time" \cite{Laine2024ATESTING}. One participant in Purovesi's \cite{Purovesi2024TestAI} raised concerns about AI doing the test planning: \textit{"Does it also then also reduce the skill of thinking, for example, when doing testing, and you ask AI to make a test plan for you, so can you then after a while do it anymore [by yourself]? I think if we say that we are experts in  testing, then we should not lose those basic skills by having someone that makes it easier for us."} \cite{Purovesi2024TestAI}. 

In the study by Ahven \cite{Ahven2022UtilizationAssurance} one participant identified potential in using AI to generate mandatory documentation required by standards, especially in security-critical domain. According to the participant, this kind of routine work does not bring value to the customer or the organization, but is still a mandatory task \cite{Ahven2022UtilizationAssurance}. Automating this kind of work with AI would have the potential of making the work more meaningful, when QA specialists would be able to focus on their core tasks, instead of routine documentation. On the other hand, when working in the security-critical domain, you  very likely do not want to use tools such as public LLMs for document generation, since then your data will be leaked.

Based on comparing the actual use cases and expected use cases, it seems that AI is underused in document generation. Modern AI tools (generative AI, LLMs) would be well suited for these use cases, and there would be great potential in making documentation related processes more efficient. However, documents should be generated only when needed, using AI to generate documentation no-one will read does not increase the overall efficiency. One of the most probable reason, why AI is underutilized in document generation are the privacy and security issues with LLMs. This would require software development organizations to invest in internal AI systems, to avoid data leaks.

\begin{table*}[!htbp]%
\centering %
\caption{Expectations vs reality - Use cases \label{table_use_case_expectations_reality}}%
\footnotesize\begin{tabular*}{\linewidth}{@{\extracolsep\fill}p{0.45\linewidth}p{0.45\linewidth}@{\extracolsep\fill}}
\toprule
\textbf{Actual Use Cases} & \textbf{Expected Use Cases} \\
\midrule

\textbf{Generation} & \textbf{Generation} \\
\hspace{3mm}Test case generation \cite{Santos2024AreTesting, Jauhiainen2024ArtificialTesting, Laine2024ATESTING, Gutierrez2020AI-PoweredDevelopment, Ahven2022UtilizationAssurance, Amalfitano2024AIPractitioners} 
& 
\hspace{3mm}Test case generation \cite{Santos2024AreTesting, Adu2024ArtificialTechniques, Ahven2022UtilizationAssurance, Jauhiainen2024ArtificialTesting, Purovesi2024TestAI, Laine2024ATESTING, Layman2024GenerativeTesting}  \\

\hspace{3mm}Code generation \cite{Adu2024ArtificialTechniques, Laine2024ATESTING, Santos2024AreTesting, Purovesi2024TestAI, Jauhiainen2024ArtificialTesting} 
& 
\hspace{3mm}Code generation \cite{Purovesi2024TestAI, Adu2024ArtificialTechniques, Layman2024GenerativeTesting, Jauhiainen2024ArtificialTesting, Laine2024ATESTING}  \\

\hspace{3mm}Test data generation \cite{Santos2024AreTesting, Jauhiainen2024ArtificialTesting, Layman2024GenerativeTesting} & 
\hspace{3mm}Test data generation \cite{Purovesi2024TestAI, Jauhiainen2024ArtificialTesting} \\


\hspace{3mm}Document generation \cite{Laine2024ATESTING, Jauhiainen2024ArtificialTesting, Santos2024AreTesting} 
& 
\hspace{3mm}Document generation \cite{Layman2024GenerativeTesting, Ahven2022UtilizationAssurance, Amalfitano2024AIPractitioners, Santos2024AreTesting, Laine2024ATESTING, Purovesi2024TestAI}  \\

\hline 
\textbf{Analysis} & \textbf{Analysis} \\
\hspace{3mm}Code and root cause analysis \cite{Adu2024ArtificialTechniques, Laine2024ATESTING, Purovesi2024TestAI, Santos2024AreTesting} 
& 
\hspace{3mm}Code and root cause analysis \cite{Purovesi2024TestAI, Jauhiainen2024ArtificialTesting, Layman2024GenerativeTesting, Laine2024ATESTING}  \\

 \hspace{3mm}Data (e.g. document, log) analysis \cite{Amalfitano2024AIPractitioners, Ahven2022UtilizationAssurance, Laine2024ATESTING} & 
 \hspace{3mm}Data (e.g. document, log) analysis \cite{Ahven2022UtilizationAssurance, Laine2024ATESTING, Purovesi2024TestAI, Layman2024GenerativeTesting, Khan2024AI-BasedTesting} \\
\hspace{3mm}Requirements analysis \cite{Santos2024AreTesting} & 
\hspace{3mm}Requirements analysis \cite{Santos2024AreTesting, Layman2024GenerativeTesting, Ahven2022UtilizationAssurance}  \\
\hspace{3mm}- & \hspace{3mm}Effort estimation \cite{Laine2024ATESTING}  \\

\hline 
\textbf{Core Testing Activities} & \textbf{Core Testing Activities}\\
\hspace{3mm}Intelligent test automation  \cite{Ahven2022UtilizationAssurance} & \hspace{3mm}Intelligent test automation \cite{Ahven2022UtilizationAssurance, Adu2024ArtificialTechniques, Amalfitano2024AIPractitioners, Amarasekara2023ChallengesProcess, Purovesi2024TestAI}  \\

\hspace{3mm}Test oracle \cite{Amalfitano2024AIPractitioners} 
& 
\hspace{3mm}Test oracle \cite{Amalfitano2024AIPractitioners} \\

\hspace{3mm}UI testing \cite{Ahven2022UtilizationAssurance, Amalfitano2024AIPractitioners} & \hspace{3mm}UI testing \cite{Purovesi2024TestAI, Ahven2022UtilizationAssurance, Amalfitano2024AIPractitioners}  \\

\hspace{3mm}- 
& \hspace{3mm}Exploratory testing \cite{Laine2024ATESTING, Purovesi2024TestAI}  \\

\hspace{3mm}- & \hspace{3mm}Security, reliability, availability, and failover testing \cite{Layman2024GenerativeTesting}  \\
\hspace{3mm}- & \hspace{3mm}Vulnerability detection \cite{Jauhiainen2024ArtificialTesting}  \\

\hline 
\textbf{Prioritization} & \textbf{Prioritization} \\
\hspace{3mm}Test case prioritization \cite{Ahven2022UtilizationAssurance, Gutierrez2020AI-PoweredDevelopment}
& 
\hspace{3mm}Test case prioritization \cite{Ahven2022UtilizationAssurance, Purovesi2024TestAI, Layman2024GenerativeTesting}  \\

\hspace{3mm}Defect prediction \cite{Jauhiainen2024ArtificialTesting}  
& 
\hspace{3mm}Defect prediction  \cite{Jauhiainen2024ArtificialTesting}  \\

\hspace{3mm}- & \hspace{3mm}Test coverage prediction \cite{Jauhiainen2024ArtificialTesting}  \\

\hspace{3mm}- & \hspace{3mm}Defect classification \cite{Jauhiainen2024ArtificialTesting}  \\

\hline
\textbf{Repair} & \textbf{Repair}\\
\hspace{3mm}Fixing bugs \cite{Santos2024AreTesting} & \hspace{3mm}Fixing bugs \cite{Layman2024GenerativeTesting, Purovesi2024TestAI, Santos2024AreTesting}  \\
\hspace{3mm}Bug reporting \cite{Ahven2022UtilizationAssurance} & \hspace{3mm}Bug reporting \cite{Santos2024AreTesting, Amarasekara2023ChallengesProcess, Laine2024ATESTING} \\

\hline
\textbf{Test Maintenance and Infrastructure} & \textbf{Maintenance and Infrastructure}\\
\hspace{3mm}Test execution \cite{Santos2024AreTesting} 
& \hspace{3mm}Test execution \cite{Jauhiainen2024ArtificialTesting}  \\
\hspace{3mm}Test maintenance \cite{Amalfitano2024AIPractitioners} & \hspace{3mm}Test maintenance \cite{Purovesi2024TestAI}  \\
\hspace{3mm}Test configuration management \cite{Ahven2022UtilizationAssurance} & \hspace{3mm}Test configuration management \cite{Laine2024ATESTING, Khan2024AI-BasedTesting}  \\
\hspace{3mm}Test optimization \cite{Amalfitano2023ArtificialStudy} & \hspace{3mm}Test optimization \cite{Laine2024ATESTING, Amalfitano2024AIPractitioners, Ramchand2022RoleAssurance}  \\
\hspace{3mm}- & \hspace{3mm}Automating the release process \cite{Santos2024AreTesting} \\

\bottomrule
\end{tabular*}

\end{table*}

\textbf{Analysis.} In the category "Analysis", we have use cases related to data analysis. A popular code related use case was "code and root cause analysis". This meant, for example analyzing what the code does, understanding legacy code, finding defects in code, or troubleshooting via root cause analysis. An interviewee in Purovesi's \cite{Purovesi2024TestAI} described the activity as follows: 
\textit{"If you think about the whole test automation aspect, it has just gone to the direction where I'm asking the AI for code scripts or things like that. And help with finding bugs - I'll enter the code snippets [to AI] and ask: "what's wrong with this?"}.

The data analysis category combined various data analysis task with various goals or purposes. System-wide activities, such as analyzing large quantities, of data, such as logs, test reports, and other historical data were mentioned. Data analysis itself is a more abstract concept, and was used to fulfill many different goals and use cases, for example, effort estimation. Laine \cite{Laine2024ATESTING} saw utilizing LLMs or machine in effort estimation would make the effort more data-driven, based on the historical data on effort estimations and actual realized efforts. The data analysis category is also closely related to to document generation, such generating summaries. On the individual level, testing involves having to read and analyze long documents, and this manual work is something that AI could help in by summarizing relevant information from the documents \cite{Laine2024ATESTING}.

Understanding the requirements is essential for a tester to perform their work well. There were no specific details available about the actual use case, how AI is utilized in requirements analysis, only survey results confirming that it was used \cite{Santos2024AreTesting}. The lack of use cases describing requirements analysis might be due to our papers focusing more on software testing. Doing a requirements analysis specific literature review might reveal more information. In Santos et al's \cite{Santos2024AreTesting} the respondents saw potential in utilizing AI for requirements analysis mostly on the individual-level, for example in extracting key information from the requirements. A system-wide vision of AI-based requirements analysis was something described by an interviewee in Ahven's \cite{Ahven2022UtilizationAssurance} study, who hoped that an AI would analyze the requirements and check based on them that the system under test (SUT) fulfills the requirements. \cite{Ahven2022UtilizationAssurance}.

\textbf{Core Testing Activities.} The overall impression in the studies was that test automation would greatly benefit from the application of AI. In an effort to distinguish, what this would mean in practice, we identified five different ways  from the studies how AI could be applied in test automation or what kind problems in test automation it could solve. They were combined them under the label of "intelligent test automation". There were surprisingly many different ways, how respondents in the studies wanted to apply AI to test automation, and what is presented here may just be a subset of the full potential. Intelligent test automation in the context of this study can be viewed as: 
\begin{itemize}
    \item A solution to test automation maintenance and configuration management \cite{Ahven2022UtilizationAssurance, Amalfitano2024AIPractitioners}: Self-healing test automation should adapt to external changes (that do not change the functionality of the the SUT), such as changes in configurations, infrastructure, library updates, refactoring. 
    \item A solution to test case maintenance \cite{Purovesi2024TestAI, Amarasekara2023ChallengesProcess, Amalfitano2024AIPractitioners}: Self-healing test automation should rewrite test cases based on the new implementation or changed requirements. 
    \item A solution to fixing bugs in SUT found by test automation  \cite{Amalfitano2024AIPractitioners}: When test automation finds new bugs, it would automatically repair them in the SUT (automated program repair).
    \item A solution to delayed test results  \cite{Ahven2022UtilizationAssurance}: predictive test automation that would analyze changes and prioritize tests based, for example, on the changes made to SUT.
    \item A solution to automatically generating new tests for new features in the SUT \cite{Purovesi2024TestAI}: test automation should create test cases based on the new implementation/requirements. 
    \item A solution to regression testing \cite{Ahven2022UtilizationAssurance}: Test automation finds errors via comparing old and new implementations. 
\end{itemize}

Intelligent test automation is dependent on test oracles. Test oracles are used to assess whether a software system performs as intended, aka is the behavior of the SUT correct or incorrect \cite{Khan2024AI-BasedTesting}. Barr et al \cite{Barr2015TheSurvey} describe "test oracle problem" as being a bottleneck that prevents greater overall test automation.  Without a test oracle, a human  has to determine whether observed behavior is correct or not \cite{Barr2015TheSurvey}. It is sometimes difficult even for humans, who are aware of the business-context and requirements, to decide, whether the behavior is correct, and should a a test or the implementation be fixed.

Amalfitano et al \cite{Amalfitano2024AIPractitioners} surveyed the use of AI in test oracle definition in GUI context, and found that several different AI approaches, such as "Knowledge Representation and Reasoning and Regression", "Text Mining", "Ontology", and "Image Processing", were used in test oracle definition, but machine learning techniques were less common. In the same study, respondents also saw potential in employing LLMs like ChatGPT top to analyze documentation and specifications for generating test oracles \cite{Amalfitano2024AIPractitioners}. Amalfitano et al \cite{Amalfitano2024AIPractitioners} found that AI can support defining the expected behaviors and outcomes of the GUI, for example by training AI models, to recognize the expected GUI responses at runtime, or to compare the GUI behaviors against learned patterns \cite{Amalfitano2024AIPractitioners}. Their results indicate that AI was helpful in finding deviations from expected behaviors, but human involvement was still needed in training AI, establishing initial parameters, and interpreting the results \cite{Amalfitano2024AIPractitioners}. 

UI testing was viewed as an expensive and high-effort activity, especially because the automated UI tests required so much continuous maintenance, and AI assistance was considered as highly welcome \cite{Ahven2022UtilizationAssurance}. UI testers were already using various AI-based tools, on system-wide and individual-level, such as Applitools, Test.ai, AutoBlackTest, UserZoom, testRigor, as well as LLMs \cite{Amalfitano2024AIPractitioners, Ahven2022UtilizationAssurance}. It seems that for UI testing, there was a variety out-of-the-box AI tools available already. It seemed that sometimes the commercial tools had limitations according to testers, such as overall usability, or the tool has been trained with applications in a specific domain, and therefore failed in testing applications in other domains \cite{Ahven2022UtilizationAssurance}.

In addition to commercial AI testing tools, also context-specific POCs had been created for UI testing. In the study by Ahven \cite{Ahven2022UtilizationAssurance}, an interviewee described a system where UI traverses the UI-tree and if errors are encountered via machine vision, it is rewarded via reinforced learning. The system was built to handle a sngle specific UI, but they had plans for expanding to other UIs as well \cite{Ahven2022UtilizationAssurance}. They felt that the results were promising \cite{Ahven2022UtilizationAssurance}.

Exploratory testing is an activity that requires high skills and experience. Even though it is usually a manual testing activity, the purpose of exploratory testing is not just to follow predetermined test cases, but to use creativity to investigate the SUT. It was indicated that AI, for example LLMs, could help in generating ideas for exploratory testing scenarios \cite{Purovesi2024TestAI, Laine2024ATESTING}. However, this would probably require, that the AI would have enough contextual and domain knowledge about the SUT. Exploratory testing, due to it's creative nature, is one of the activities, that is not possible to automate with current AI solutions. 

Although no actual use cases were mentioned, potential was also seen for AI-supported or completely AI-driven vulnerability detection \cite{Jauhiainen2024ArtificialTesting}. Similarly topics such as security, reliability, availability, and failover testing were mentioned as potential venues for practical trials and implementation projects. \cite{Layman2024GenerativeTesting}.

\textbf{Prioritization. }  AI-driven test case prioritization were seen as solutions to delayed test results in situations where automated tests runs may last for hours, even full days. The expected benefit that would result from test case prioritization in test automation is time savings, when potential defects are found earlier. According to Khan et al\cite{Khan2024AI-BasedTesting}, AI-based algorithms can be utilized in ranking test cases on their ability of finding flaws, reducing testing time and effort.  

AI could potentially do prioritization also be based on test coverage and critical parts of the software:

\textit{"We would be able to easily get some kind of analyses that how big part of our application we have now tested or where our focus is on testing. Or learn, for example, if we have a web application, that what is the human behavior on the page, that we will find those important critical parts of what needs to test more."} \cite{Purovesi2024TestAI}

POC's on test case prioritization were also done, and interviewee from Ahven's \cite{Ahven2022UtilizationAssurance} described a POS that prioritized test cases with artificial intelligence. It was based on collecting historical data from precious test runs, and the goal was that the AI could select the most important tests to run to get feedback faster \cite{Ahven2022UtilizationAssurance}.

One person mentioned defect prediction in Jauhiainen's \cite{Jauhiainen2024ArtificialTesting} study, when asked how they, or the customer they are working for, are using AI in testing. Unfortunately, there was no additional details about this use case available. Test coverage prediction and defect classification were also mentioned in the same study, as something should AI help in or solve, but no more details about the expectations were available \cite{Jauhiainen2024ArtificialTesting}.

\textbf{Repair.} The items in the category of "Repair", can be understood as software development activities, rather than traditional software testing activities. Maintaining test automation often involves bug fixing. And if we take into account the shift-left paradigm, where testing and quality is included from the beginning of the project, not as the last step, and also, intelligent automation that fixes bugs and defects automatically, the border between testing and development activities gets blurred.

One respondent in Santos et al's \cite{Santos2024AreTesting}, indicated that they have used LLMs in bug fixing on individual-level. They did not specify, what was the context of bug fixing, SUT or, for example, the test infrastructure side. Similarly, the expectations were quite general-level observations, such as the one from the virtual round-table by Layman and Vetter \cite{Layman2024GenerativeTesting}:

\textit{"AI may also offer trustworthy guidance to stakeholders on the documentation, prioritization, and even repair of defects, and potentially the release-readiness of whole systems."}

In Santos et al's  \cite{Santos2024AreTesting} study, the second most anticipated use of LLMs among non-users was bug reporting activities, while current users expected using LLMs in bug fixing support. Amasekara et al \cite{Amarasekara2023ChallengesProcess} also saw potential in using ML tools for reporting bugs automatically in bug tracking tools, as did Laine \cite{Laine2024ATESTING}, indicating that AI could help in finding information from logs, about dependencies and connecting the report to a specific part of the system or a feature. An interviewee in Ahven's \cite{Ahven2022UtilizationAssurance} thesis described an early stage experimental project with promising initial results, where bug reports were automated using AI \cite{Ahven2022UtilizationAssurance}. When a customer reported of a problem, based on that report an automated test was generated that reproduces the problem. The developer could run the test on their machine and verify if the problem has been fixed \cite{Ahven2022UtilizationAssurance}. 

\textbf{Test Maintenance and Infrastructure.} Regarding the actual use case of test execution, there were no additional details available, only information that  specialists use AI in test execution \cite{Santos2024AreTesting}. Using AI in test execution, such as integrating AI in CI/CD -pipelines, was described in the background literature in some of the the studies, but there was only the one mention of an actual use case, and were no mentions of expectations related to it. It seems that there is a lack of empirical studies describing experiences of AI-based test execution, and overall the AI-enhanced or AI-based testing infrastructure.

Test maintenance category contains many different tasks, such as test case maintenance and test infrastructure maintenance. Overall, test maintenance can be very labor-intensive and usually has to be performed urgently. For example, changes in SUT, test frameworks, libraries, infrastructure, test code, test cases can cause the test automation to break. Paradoxically, even though test automation is often seen as a solution to minimizing human labor, or automating routine work, it might not be that in reality. An interviewee in Ahven's \cite{Ahven2022UtilizationAssurance} described the maintenance effort (translated from Finnish):

\textit{"So, typically it will take us, what did it take, we had four full time testers and half of them did test automation maintenance, which meant that we did not move forwards, just kept the test automation alive."}

From the description, "we did not move forwards, just kept the test automation alive", we can sense, that the interviewee in question would like to have spent their time doing more productive tasks, than just keeping the test automation alive. In Purovesi's \cite{Purovesi2024TestAI} thesis, multiple interviewees hoped that AI would help in test maintenance, but specifics on how this would be done in practice were not mentioned. In Amaltafino et al's \cite{Amalfitano2023ArtificialStudy} a respondent mentioned "the possibility of providing more easily maintained test suites when the AI generators are fed with the modifications applied to the SUT".

In test configuration management, or test environment management,  AI can be potentially used to manage hardware, software, firmware and other configurations \cite{Laine2024ATESTING}. In Ahven's \cite{Ahven2022UtilizationAssurance} study, an interviewee described a custom system that was developed for their customer, where thousands of test cases were run on physical devices. Before the AI-based solution, someone had to manually assign the tests that should be performed in the specific devices \cite{Ahven2022UtilizationAssurance}. With AI, the system could assign the the correct tests to the correct devices automatically, system adapts automatically when devices are changed, and manual testing of devices is not longer needed, resulting in reduced effort and time savings \cite{Ahven2022UtilizationAssurance}. Khan et al \cite{Khan2024AI-BasedTesting} suggest utilizing AI in test environment management as maintaining stable and well configured test environments is challenging, especially for complex systems. They suggest that by "autonomously configuring and controlling the test environment based on historical data and machine learning algorithms, AI-based software testing can contribute to improving test environment management" \cite{Khan2024AI-BasedTesting}.

Test optimization is closely connected to the test maintenance category, but we separated it into a separate use case containing the non-urgent tasks, such as test improvements. Not all tests are useful, and some tests overlap or have unnecessary repetition. Laine \cite{Laine2024ATESTING} identified potential use cases for optimizing the test order to avoid repetitive test initiation steps, and identifying overlapping or obsolete tests case. This would result in time savings via faster feedback cycles, as the number of tests to run and repetition would be reduced \cite{Laine2024ATESTING}. In Ramchand et al's \cite{Ramchand2022RoleAssurance} the participants were asked if AI will help reducing the number of written test cases required, and and 85 percent of the respondents agreed, while 10 percent disagreed and the remaining 5 percent ratio was not sure about the answer". This would also indicate that AI can has potential in optimizing test cases  \cite{Ramchand2022RoleAssurance}. With test optimization, one of the goals is to increase coverage \cite{Amalfitano2024AIPractitioners}. In Amaltafitano et al's \cite{Amalfitano2024AIPractitioners}, respondents indicated that LLMs could be use to "ensure multilingual test cases, or the enhancement of existing test cases to provide better conformance towards black-box testing standards.

Santos et al \cite{Santos2024AreTesting} saw also potential in using AI to automate the release process: "we hypothesize their potential use in generating user guidelines and automating the release process, including tasks involving multiple code branches". They also highlighted the need to perform additional research to investigate the possibilities of AI-automated release processes  \cite{Santos2024AreTesting}.

\subsubsection{Expectations versus Reality: Benefits}

When looking at the actual versus expected benefits (Table \ref{table_benefits}), we can see some differences between the expectations and actual benefits. Only six studies \cite{Barenkamp2020ApplicationsEngineering, Gutierrez2020AI-PoweredDevelopment, Laine2024ATESTING, Amalfitano2024AIPractitioners, Santos2024AreTesting, Ahven2022UtilizationAssurance} reported actual observed benefits, but expected benefits were mentioned in a wider number of studies.

\begin{table*}[!htbp]%
\centering %
\caption{Expectations vs reality -  Benefits} 
\label{table_benefits}%
\footnotesize\begin{tabular*}{\linewidth}{@{\extracolsep\fill}lllll@{\extracolsep\fill}}
\toprule
\textbf{Actual Benefits} & \textbf{Expected Benefits} \\
\midrule
Time savings\cite{Gutierrez2020AI-PoweredDevelopment, Laine2024ATESTING, Barenkamp2020ApplicationsEngineering, Ahven2022UtilizationAssurance} 
& Time savings \cite{Amarasekara2023ChallengesProcess, Khan2024AI-BasedTesting, Purovesi2024TestAI, Barenkamp2020ApplicationsEngineering, Hossain2024AICompanies, Ahven2022UtilizationAssurance} \\

Better coverage \cite{Amalfitano2024AIPractitioners, Gutierrez2020AI-PoweredDevelopment, Santos2024AreTesting} & Better coverage \cite{Amarasekara2023ChallengesProcess, Purovesi2024TestAI, Khan2024AI-BasedTesting, Adu2024ArtificialTechniques} \\

Better resource allocation \cite{Gutierrez2020AI-PoweredDevelopment, Ahven2022UtilizationAssurance} & Better resource allocation\cite{Purovesi2024TestAI} \\

Increased productivity and efficiency \cite{Laine2024ATESTING, Amalfitano2024AIPractitioners, Jauhiainen2024ArtificialTesting, Purovesi2024TestAI} & Increased productivity and efficiency \cite{Laine2024ATESTING, Barenkamp2020ApplicationsEngineering, Khan2024AI-BasedTesting, Purovesi2024TestAI, Gutierrez2020AI-PoweredDevelopment, Amarasekara2023ChallengesProcess} \\

Quality improvement \cite{Gutierrez2020AI-PoweredDevelopment} & Quality improvement \cite{Layman2024GenerativeTesting, Adu2024ArtificialTechniques, Laine2024ATESTING, Purovesi2024TestAI, Barenkamp2020ApplicationsEngineering, Hossain2024AICompanies} \\

Increased accuracy and precision \cite{Amalfitano2024AIPractitioners}  & Increased accuracy and precision \cite{Khan2024AI-BasedTesting, Hossain2024AICompanies, Adu2024ArtificialTechniques, Purovesi2024TestAI}  \\

AI will make testing more accessible for everyone \cite{Amalfitano2024AIPractitioners} & AI will make testing more accessible for everyone \cite{Layman2024GenerativeTesting, Amarasekara2023ChallengesProcess, Laine2024ATESTING, Amalfitano2024AIPractitioners}  \\

-  & Cost savings \cite{Barenkamp2020ApplicationsEngineering, Purovesi2024TestAI, Bhuvana2023IntegrationDevelopment, Laine2024ATESTING} \\

- & Increased job satisfaction \cite{Laine2024ATESTING, Ramchand2022RoleAssurance, Hossain2024AICompanies}  \\
- & Improved communication \cite{Purovesi2024TestAI, Barenkamp2020ApplicationsEngineering, Bhuvana2023IntegrationDevelopment} \\
\bottomrule
\end{tabular*}
\end{table*}

A lot of expectations are placed on AI in software testing. An interviewee from  Purovesi's \cite{Purovesi2024TestAI} study summarized, what kind of benefits they expect from AI especially in the test automation context:

\textit{"You get more efficient and effective testing, you get more out of less, you get it done faster. Less human resources needed for maintenance, better test coverage. Faster, in a shorter time cycle, the whole testing process from design to reporting and fewer bugs." }

Similar sentiment was expressed in the study by Bhuvana et al \cite{Bhuvana2023IntegrationDevelopment}:

\textit{"Artificial intelligence improves the process of identifying software bugs, anticipates the results of tests, and automates the testing of software, resulting in cost reduction and making the work of software developers easier."}

To really drill down to the concrete benefits observed in the studies regarding AI adoption in software testing, we further classified the benefits in the studies to three categories: benefits from AI in general, technologies and use cases. In the "benefits from AI in general" category, we found high level blanket statements in the style of "AI in testing brings cost savings" or "AI based tools help increase productivity". In the "benefits from technologies" category we found slightly more concrete statements, such as "Deep learning algorithms speed up processes". The "benefits from use cases" category was the most interesting to use, since it provided the most concrete information on how benefits are achieved in practice. Both actual and expected benefits were reported in these three categories. 

We decided to explore the category "benefits from use cases" in more detail. In the table \ref{table_use_cases_and_benefits}, we mapped the connections between use cases and benefits. The table contains the use cases that were explicitly linked to specific benefits. Explicit linking means statements such as "AI-based test case generation increased test coverage". Actual-column indicates, that there was an actual use case, and there was an actual benefit reported related to the use case. Expectation-column indicates, that there is an link between expected benefit and expected use case, or that there is an actual use case with expected benefits. 

\begin{table*}[!htbp]%
    \caption{Use cases (actual and expected) explicitly associated with benefits } \label{table_use_cases_and_benefits}
    \centering %
    \footnotesize\begin{tabular*}{\linewidth}
        {llll}
        \toprule
        \textbf{Benefit} & \textbf{AI-Assisted Use Cases} & \textbf{Actual} & \textbf{Expectation}\\
        \midrule
         Time savings & Test case generation & x \cite{Gutierrez2020AI-PoweredDevelopment} & x \cite{Adu2024ArtificialTechniques, Gutierrez2020AI-PoweredDevelopment} \\
            & Code generation  & x \cite{Gutierrez2020AI-PoweredDevelopment, Santos2024AreTesting} & \\
            & Code and root cause analysis  & x \cite{Barenkamp2020ApplicationsEngineering} &\\
            & Intelligent test automation & x \cite{Ahven2022UtilizationAssurance} & x \cite{Amarasekara2023ChallengesProcess} \\
            & Test case prioritization & x \cite{Ahven2022UtilizationAssurance} & x \cite{Khan2024AI-BasedTesting}\\
            & Defect prediction  &  x \cite{Barenkamp2020ApplicationsEngineering, Gutierrez2020AI-PoweredDevelopment} & \\
            \hline
        Better coverage & Test case generation  & x \cite{Amalfitano2024AIPractitioners, Gutierrez2020AI-PoweredDevelopment, Santos2024AreTesting} & x \cite{Layman2024GenerativeTesting, Adu2024ArtificialTechniques} \\
            & Test data generation  & x \cite{Santos2024AreTesting} & \\
            & Data analysis  & & x \cite{Khan2024AI-BasedTesting} \\
            & Intelligent test automation &  & x \cite{Amarasekara2023ChallengesProcess} \\
            & Test case prioritization &  & x \cite{Khan2024AI-BasedTesting}\\
            & Defect prediction & & x\cite{Gutierrez2020AI-PoweredDevelopment}  \\
            & Test optimization & & x\cite{Amalfitano2023ArtificialStudy}  \\
            \hline         
        Better resource allocation & Test case prioritization  & & \cite{Gutierrez2020AI-PoweredDevelopment} \\
            & Intelligent test automation & x \cite{Ahven2022UtilizationAssurance, Amalfitano2024AIPractitioners} & x \cite{Amarasekara2023ChallengesProcess}\\
            & Test configuration management  & x \cite{Ahven2022UtilizationAssurance} &\\
            & Bug reporting &  & x \cite{Amarasekara2023ChallengesProcess}\\
            & Defect prediction & & x \cite{Gutierrez2020AI-PoweredDevelopment}  \\
            \hline
        Increased productivity and efficiency & Code generation  & x \cite{Laine2024ATESTING} & \\
            & Test case generation & x \cite{Amalfitano2024AIPractitioners} & x \cite{Adu2024ArtificialTechniques} \\
            & Defect prediction & & x \cite{Gutierrez2020AI-PoweredDevelopment}  \\
            & Intelligent test automation  & & x \cite{Amarasekara2023ChallengesProcess, Adu2024ArtificialTechniques}\\
            \hline
        Quality improvement & Test case generation  & x \cite{Gutierrez2020AI-PoweredDevelopment} & \\
            & Code generation  &  & x \cite{Layman2024GenerativeTesting}\\
            \hline   
        Increased accuracy and precision & -  &   & \\ 
            \hline
        AI will make testing more accessible for everyone & Intelligent test automation & & x \cite{Amarasekara2023ChallengesProcess} \\
            & Code generation  &  & x \cite{Laine2024ATESTING} \\
             \hline
        Cost savings & Intelligent test automation & & x \cite{Bhuvana2023IntegrationDevelopment} \\
            \hline
        Increased job satisfaction & Test case generation & & x \cite{Khan2024AI-BasedTesting}\\ 
            & Document generation & & x \cite{Bhuvana2023IntegrationDevelopment}  \\
            & Code generation  &  & x \cite{Laine2024ATESTING} \\
            \hline
        Improved communication & - & & \\
        
        \bottomrule
    \end{tabular*}
\end{table*}

\textbf{Time savings} was a category where we grouped together multiple different items related to time. The actual observed benefits (see table \ref{table_benefits}) contained shorter and/or faster testing process \cite{Gutierrez2020AI-PoweredDevelopment}, faster development \cite{Gutierrez2020AI-PoweredDevelopment}, bugs are found earlier \cite{Barenkamp2020ApplicationsEngineering}, faster troubleshooting \cite{Laine2024ATESTING} and faster feedback cycles \cite{Ahven2022UtilizationAssurance}. The expectations related to time savings in table \ref{table_benefits} were shorter and/or faster testing process\cite{Amarasekara2023ChallengesProcess, Khan2024AI-BasedTesting, Purovesi2024TestAI}, faster development \cite{Barenkamp2020ApplicationsEngineering}, bugs are found earlier \cite{Purovesi2024TestAI, Amarasekara2023ChallengesProcess}, faster feedback cycles \cite{Ahven2022UtilizationAssurance}, faster release process \cite{Amarasekara2023ChallengesProcess} and faster time-to-market \cite{Purovesi2024TestAI, Hossain2024AICompanies, Gutierrez2020AI-PoweredDevelopment, Hossain2022ApplicationSurvey}.

Actual time savings were observed  as a result of multiple generation, analysis, intelligent test automation and prioritization activities (see table \ref{table_use_cases_and_benefits}). When comparing tables \ref{table_benefits} and \ref{table_use_cases_and_benefits}, we can see that not all the expectations were mapped explicitly to specific use cases. Instead, time savings was reported more as general expectation, as a benefit from AI in general or as a benefit from AI technologies. 

Time savings seemed to be the most prevalent type of actual benefit that could be achieved via AI adoption. However, there was also evidence of AI not bringing time savings, two different interviewee from Purovesi's \cite{Purovesi2024TestAI} thesis described the situation, especially in the context of test automation:

\textit{"It depends so much on whether you are using the own instance where you can use your own data. I don't think it really speeds up that much yet."} 

\textit{"I don't think it saves time, it seems to me that it takes more time than it saves, as it's really only now being studied and built.} 

\textbf{Better coverage} meant in general achieving a higher test coverage than via traditional means, usually with less effort. It had been observed as a result of test case and test data generation use cases. Better coverage as a concept is one of benefits that is quite easy to measure and observe in practice, especially in the case of white-box testing. 

\textbf{Better resource allocation} meant that less manual labor, or human resources in general, were needed, or that the management of technical resources, such as devices, was more efficient. The especially in the expectation side, focus was clearly on the allocation of human resources, and for example, better overall allocation of computing resources was not mentioned as such. Better resource allocation was observed as result of intelligent test automation and test configuration management. 

\textbf{Increased productivity and efficiency} had been explicitly only observed as a result of code generation and test case generation in the context of the studies we analyzed (see table \ref{table_use_cases_and_benefits}). However, in some cases the respondents felt that the efficiency improved only a little \cite{Jauhiainen2024ArtificialTesting, Purovesi2024TestAI}. Test case generation, defect prediction and intelligent test automation were expected to bring improvements in productivity and efficiency. Again, in several studies it was expected that AI in general or AI technologies would increase productivity and efficiency, but the concrete use cases (or other details) on how the benefit could be realized were missing. 

\textbf{Quality improvement} contained observations such as less defects or bugs, they are detected earlier, or in general higher quality. Quality improvement was explicitly reported as a result of test case generation, and was expected as a result of code generation. Quality improvement itself is such a self-evident result of software testing in general, so even though it was not mentioned directly, it is the main driver for software testing. Studies where quality improvement achieved by AI is measured in real-life settings would be interesting to see.

\textbf{Increased accuracy and precision} was not directly connected to any use cases, but Amaltafitano et al \cite{Amalfitano2024AIPractitioners} had observed that according to their respondents, AI was seen "as a valuable help in automated testing as it increases testing accuracy and precision and uncovers blind spots in GUI-based testing". However, AI hallucinations were seen as a potential threat to the accuracy \cite{Purovesi2024TestAI}. Unlike for example, better coverage, increased accuracy and precision as a concept is slightly more vague and more difficult to measure. Investigation into the meaning and possible measures of accuracy and precision in the AI context would be needed, especially to see how potential hallucinations affect it.

Especially due to the new AI tools and technologies (LLMs, genAI), it was expected that \textbf{AI will make testing more accessible for everyone}. Previously, AI utilization in software testing has required a highly specialized skill-set, and deep knowledge about AI technologies, but now it is more accessible to even for those who do not have deep technical understanding of AI. The modern AI tools offer a lower barrier to entry or a threshold for application and implementation  \cite{Laine2024ATESTING, Purovesi2024TestAI}. In addition, Amaltafitano et al  \cite{Amalfitano2024AIPractitioners} found that AI has a potential of easing the work of inexperienced testers. On the other hand, an interviewee in Purovesi's \cite{Purovesi2024TestAI} study questioned whether junior testers have the capability of critically evaluate AI results, as in can they tell if the results are good enough or they just automatically accept what AI has produced. 

Thomas Dohmke saw in the virtual round-table by Layman and Vetter \cite{Layman2024GenerativeTesting} that the accessibility may have also greater implications in the context of software development and society:

\textit{"As such, AI will democratize access to software development and will significantly increase the number of people that have the skills to accelerate human progress." }

Interestingly, \textbf{cost savings} had not been reported in connection to AI adoption. It seems that in the short term, especially in the case of early adoption phase, AI adoption in software testing does not bring cost savings, but instead increases costs due to the investments required. Investments are required at least in infrastructure, technologies, resourcing and training of personnel. A lack of industry standards and guidelines means that companies had to start their investigation often from scratch, as there was a lack of reference implementations. On the other hand, potential reasons why cost savings were not mentioned related to actual use cases, was that it was not specifically covered in the studies, or actively measured, or that the participants of the studies did not have access to financial data. It was noted that it was still difficult to evaluate the return-of-investment (ROI) of AI projects \cite{Baqar2024TheValidation}.

\textbf{Increased job satisfaction} was something that was expected as a result of AI replacing repetitive, mundane and tedious tasks \cite{Laine2024ATESTING, Ramchand2022RoleAssurance}. This would have the potential for freeing the testing experts' time for higher level decision-making, making their work more meaningful and interesting, in order to focus on more creative and strategic tasks \cite{Laine2024ATESTING, Hossain2024AICompanies}.

Hossain et al \cite{Hossain2022ApplicationSurvey} found in their survey, that poor communication in teams involved in software testing can lead to delays, mistakes, and misunderstandings. One of the potential benefits of AI could be \textbf{improved communication}. In Purovesi's \cite{Purovesi2024TestAI} study, AI was seen as a potential driver for better communication, by getting rid of organizational silo's that cut off information sharing \cite{Purovesi2024TestAI}. This was seen as a wider change, not just impacting testing \cite{Purovesi2024TestAI}. Similar opinions were expressed by a participant in Barenkamp et al's \cite{Barenkamp2020ApplicationsEngineering} study in a wider software development scope, they felt that "AI improves the efficiency of software delivery processes, it eases team collaboration and the integration of customer feedback in code". Bhuvana et al \cite{Bhuvana2023IntegrationDevelopment} also indicated that AI could aid in the collaboration between developers and clients in the planning phase of software projects. There would be potential for researching how AI could be used to improve communication and collaboration in software testing and development, or investigating how AI affects communication and information sharing in the software testing and development contexts.

\subsubsection{The Scope of AI Adoption}

In the details of the use case descriptions in several studies, we noticed a common pattern on the scope of the adoption: the use cases for AI in software testing could be individual-level or system-wide activities. In the individual-level AI, especially LLMs, could be described as more of a personal assistant  or a digital buddy to the testers \cite{Purovesi2024TestAI, Barenkamp2020ApplicationsEngineering}. 

\textit{"I see its potential everywhere, it speeds up the start, or whatever phase is ongoing, if it involves any kind of writing or other kind of manual work. To speed up and of course it can also explore, perhaps getting quick answers if you feel that you have some problems in any of your work, then maybe it can give you some ideas, when you are blinded to what you are doing. That it can find relevant points then. Kind of like a "buddy", with whom to go through things."} \cite{Purovesi2024TestAI}

The system-wide adoption meant that the AI adoption is done on a higher level in the organization, and it has a larger impact, for example eliminating some manual tasks completely. 

If we take test case generation as an example, individual-level use case could mean that the specialist generates test cases with an LLM for the specific feature they are testing manually. An example of test case generation as a system-wide use case then would be when test cases are automatically generated by AI-enhanced test automation for the new features.

In table \ref{tab:use_cases_system_individual} we mapped the identified use cases to the scope of the use case. The scope of the use cases displays AI adoption in testing is done on different levels in the studies we analyzed. In essence it means, whether the use case was described to be performed by an individual or to be implemented as a system-wide automated process. The x mark indicates that we could determine if the use case was system-wide or individual level. And - mark indicates that we could not determine the scope of the use case in the studies. 

\begin{table*}[!htbp]%
    \caption{The scope of the use cases: system-wide and individual level use cases based on the descriptions in the studies. x = the scope was mentioned, -  = no observation or inconclusive}
    \label{tab:use_cases_system_individual}
    \centering
    \footnotesize\begin{tabular*}{\linewidth}
    {lcc}
        \toprule
        \textbf{Use case} &  \textbf{System-Wide} & \textbf{Individual-level} \\
        \midrule
        \textbf{Generation} & & \\
        \hspace{3mm}Test case generation & x & x  \\
        \hspace{3mm}Code generation & - & x \\
        \hspace{3mm}Test data generation & - & - \\
        \hspace{3mm}Document generation & - & x \\
        \hline 
        \textbf{Analysis} & & \\
        \hspace{3mm}Code and root cause analysis & - & x \\
        \hspace{3mm}Data analysis & x & x \\
        \hspace{3mm}Requirements analysis & x & x \\
        \hspace{3mm}Effort estimation & x & - \\
        \hline 
        \textbf{Core Testing Activities} & & \\
        \hspace{3mm}Intelligent test automation & x & - \\
        \hspace{3mm}Test oracle & x & - \\
        \hspace{3mm}UI testing & x & x \\
        \hspace{3mm}Exploratory testing  & - & x \\
        \hspace{3mm}Security, reliability, availability, and failover testing  & - & - \\
        \hspace{3mm}Vulnerability detection  & x & - \\
        \hline 
        \textbf{Prioritization} & & \\
        \hspace{3mm}Test case prioritization & x & - \\
        \hspace{3mm}Defect prediction & - & - \\
        \hspace{3mm}Test coverage prediction  & x  & - \\
        \hspace{3mm}Defect classification   & - & - \\
        \hline 
        \textbf{Repair} & & \\
        \hspace{3mm}Fixing bugs & x & x \\
        \hspace{3mm}Bug reporting & x & - \\
        \hline 
        \textbf{Test Maintenance and Infrastructure} & & \\
        \hspace{3mm}Test execution & - & - \\
        \hspace{3mm}Test configuration management & x & - \\
        \hspace{3mm}Test optimization & - & - \\
        \hspace{3mm}Test maintenance & x & - \\
        \bottomrule
    \end{tabular*}
\end{table*}

In general, most of the use cases could be utilized or implemented on both levels, but it should be a conscious decision, something that should be considered in the organization's AI strategy. For example, should the focus be on boosting the individual-level productivity via AI-based test case generation, or boosting the system-wide productivity via AI-based test case generation, or a combination of both? 

When making the decision, organizations have to take into account, that different skill-sets, different processes and different ownership structures are needed, whether you adopt AI on a system-wide or individual-level. This affects the required investments needed.

\subsubsection{The Current State of AI Adoption in Software Testing}

To put things into perspective, we wanted to know, to what extent AI is utilized in software testing activities in the industry. The theme "The Current State of AI Adoption in Software Testing" was developed, because we needed to provide important contextual information related to our other themes. The implications and utility of evaluating the expectations vs reality would be different in a situation where AI adoption is very prevalent and when it is not.

\begin{table*}[!htbp]%
\centering %
\caption{AI adoption in software testing\label{table_ai_adoption}}%
\footnotesize\begin{tabular*}{\linewidth}{@{\extracolsep\fill}p{0.44\linewidth}p{0.02\linewidth}p{0.02\linewidth}p{0.18\linewidth}p{0.07\linewidth}l@{\extracolsep\fill}}
\toprule
\midrule
\textbf{AI in adoption in software testing} & \textbf{Sample size} & \textbf{Year} & \textbf{Description}  & \textbf{Location} & \textbf{Study}\\
33.33 percent of respondents were familiar with AI testing tools & 19 & 2022 & Survey & Bangladesh & \cite{Hossain2024AICompanies} \\\hline 
22 percent of interviewees are currently using AI tools in their work & 26 & 2024 & Survey & Finland, Poland & \cite{Jauhiainen2024ArtificialTesting} \\\hline 
48 percent of respondents had used LLMs in testing activities & 83 & 2024 & Survey & Various  & \cite{Santos2024AreTesting} \\\hline 
73.3 percent of respondents had applied AI in GUI based testing & 45 & 2024 & Survey & Various  & \cite{Amalfitano2024AIPractitioners} \\\hline 
1 interviewee out of 15 has used AI for work purposes & 15 & 2024 & Case study & Finland & \cite{Laine2024ATESTING} \\
\bottomrule
\end{tabular*}
\end{table*}

AI utilization in software testing was measured in various ways in the studies, mostly focusing on individuals' knowledge of software testing. AI adoption on organizational or company level was not measured. 

In their survey on the use of LLMs in software testing, Santos et al \cite{Santos2024AreTesting} classified the answers further based on testing activities. They found that 40 percent of professionals utilize LLMs during the initial phases of software testing (requirements analysis, test design, and test plan creation) \cite{Santos2024AreTesting}. In the intermediate stages, such as test case preparation and test execution, 28 percent of professionals used LLMs  \cite{Santos2024AreTesting}. Lastly, 32 percent of participants used LLMs in post-testing activities, for example providing support for bug fixing, regression testing, and software release  \cite{Santos2024AreTesting}. 

By just looking at the percentages, in some cases, there seeme to be quite many people adopting AI in their software testing activities, or at least being familiar with AI tools. However, the numbers do not tell all. Even though Amalfitano et al \cite{Amalfitano2024AIPractitioners} found that over 70 percent of respondents utilized AI in GUI based testing, their investigation indicated, that "GUI based testers tend to use AI-based mechanisms in a widespread yet superficial manner". 

Although many different use cases for AI-based software testing existed, it was also observed that they were not always very beneficial, or they were still on the investigation or proof-of-concept (POC) level. For example using AI in generating test data was not viewed as efficient, since AI was seen as capable of only generating basic data \cite{Purovesi2024TestAI}. People were also somewhat skeptical about new tools due to past experiences: taking new tools into use had not made work easier, because it had required a lot of additional work (such as data entry), and even at best, the amount of work stayed the same \cite{Laine2024ATESTING}. Also, learning how to use new tools is time-consuming \cite{Jauhiainen2024ArtificialTesting}. 

Another downside that was mentioned was that maintaining AI generated artifacts may be more difficult than human-generated ones. In Jauhianen's \cite{Jauhiainen2024ArtificialTesting} study one respondent noted that creating the test cases is easy, but maintenance is not: making changes to or finetuning AI-generated test cases is difficult and time-consuming \cite{Jauhiainen2024ArtificialTesting}.

Overall, among the experts, evaluating the actual benefits of AI adoption software testing was considered as difficult, or the benefits were vague. Benefits were indeed reported, but in some cases it was difficult to quantify them \cite{Purovesi2024TestAI}, or respondents felt that the efficiency improved only a little \cite{Jauhiainen2024ArtificialTesting, Purovesi2024TestAI}. Some testers felt that there was a lack of concrete estimates about the time-saving of AI assisted test automation, as the evidence was limited. \cite{Purovesi2024TestAI}. Some felt that AI utilization in testing was still in the pilot phase or POC level \cite{Purovesi2024TestAI}. Also, aggressive marketing, unrealistic promises and a "hype peak" were observed \cite{Purovesi2024TestAI}.

AI adoption in software testing is still in it's early stages. This quote from Purovesi's \cite{Purovesi2024TestAI} thesis captures the essence of the current state of AI adoption:

\textit{"Everyone is a little bit confused about what's going on. Everyone is waiting to understand, what is the best [way to utilize AI] in our context. Now is perhaps a kind of experimental phase and everyone is looking for situations where they could try [AI]." }

In the virtual round-table by Layman and Vetter \cite{Layman2024GenerativeTesting}, Adam Porter states what kind of thinking is behind the unrealistic expectations: "as with many trendy technologies (e.g., Blockchain is one recent example) there’s a real lack of understanding about what GenAI is, how it might actually be used, its benefits over existing technologies, and its potential downsides". This was seen as leading to thinking that AI can solve every problem that exists, and that the potential use cases and applications are unlimited \cite{Layman2024GenerativeTesting}. The belief that AI will solve all the problems was seen as a challenge especially when working with customers, as it was difficult to meet those unrealistic customer expectations related to AI, and get the customer to understand the true capabilities of AI \cite{Purovesi2024TestAI}

\section{Discussion}\label{sec4}

AI adoption in software testing is still an activity for the early birds. No industry-wide conventions have been established yet, which means software development organizations are left alone experimenting with AI-based solutions for testing, and bearing the cost of investigation alone. Known use cases for AI in testing do exist, but they still seem to be on the Proof-Of-Concept level, and not highly beneficial in all cases. Still, AI in software testing was seen as a potential competitive advantage: companies that do not successfully utilize AI in software testing will lose in the competition \cite{Ahven2022UtilizationAssurance}.

 Despite the lag in actual use cases and benefits, expectations regarding AI remain high and optimistic, as they have been for some time. In a survey by King et al \cite{King2019AIPerspectives}, conducted in 2017, 88 percent of the respondents expected AI to have a significant impact on automated testing between 2019 and 2025. As based on our analysis of the studies reported on this topic to the date, it seems that we are not there yet. The expectations related to the benefits of AI are still quite same as in the study by Hourani et al \cite{Hourani2019TheTesting} in 2019: software development life-cycle, and testing time, will shorten and productivity will increase. The situation is similar also in other fields. Khanfar et al \cite{Khanfar2025FactorsReview} have noticed that despite AI's potential benefits of increasing revenue, reducing costs and enhancing performance, the adoption by organizations has fallen short of expectations, leading to unsuccessful implementations. One potential way of evaluating the actual productivity and efficiency benefits of AI adoption in software testing are field experiments similar to Dell'Acqua et al's \cite{DellAcqua2023NavigatingQuality} study, where "the performance implications of AI on realistic, complex, and knowledge-intensive tasks" were evaluated.

The mismatch between expectations and true capabilities of AI has been noted in other previous research as well \cite{Kinney2024ExpectationSystems, Brennen2020WhatNews}. According to Brennen \cite{Brennen2020WhatNews}, one of the sources of unrealistic expectations is the media, where AI is viewed as being able to solve almost any problem. This can lead to too high expectations: the potential of AI is overestimated, and true costs and risks are downplayed \cite{Brennen2020WhatNews, Kinney2024ExpectationSystems}. 

From a terminological point-of-view, this study explored the utilization of AI on a fairly high abstraction level. For example, the term AI itself holds many different meanings and refers to a multitude of technologies and tools, from machine learning algorithms to LLMs. Also, the use cases and benefits were sometimes reported with a very high abstraction. Additional empirical studies would be needed, where the costs and benefits of AI adoption projects in software testing are evaluated in more detail. In this context, it would be also important to investigate different types of AI adoption projects: a project for utilizing deep learning algorithms in automated testing could be different from utilizing LLMs in the context of exploratory testing. 

According to Jensen et al \cite{Jensen2024ManagingStudy}, depending on the unique features of different software development organizations, expectations towards AI tool adoption should be evaluated case by case. There might not be a one-size-fits-all tool, but a variety of tools for different software development organizations, some becoming more mainstream than others \cite{Jensen2024ManagingStudy}.

As stated, the benefits over other technologies should be evaluated, as AI is only one potential tool in a range of testing tools, and not all problems might warrant an AI-level solutions. In Ahven's \cite{Ahven2022UtilizationAssurance} study, one interviewee predicted that the overall trend in testing, irrespective of AI, will lead to a reduction in the total time allocated for testing by developers and testers, owing to the advent of more advanced testing tools. AI might not be the optimal solution to all problems, and for example rule-based systems were still seen as a valid and potentially resource-friendly option, even though setting up such systems required effort \cite{Ahven2022UtilizationAssurance}. 

Our RQ2 was: how is AI utilized in software testing in the industry? We found that there is a multitude of actual and especially potential use cases for AI in software testing, not only in the testing activities themselves, but supporting activities such as test and document generation, analysis, test maintenance and infrastructure. The use cases mentioned in this study are perhaps just a scratch on the surface of all the potential use cases. Overall, the data available about actual use cases in the industry was not very detailed among the studies we analyzed. For example, some of the studies were surveys that did not provide explanations or details about the use case. They only confirmed that an use case exists. Each use case mentioned in the study would have a potential for a deeper investigation in the industry setting. 

Another interesting finding on this study was the scope AI use cases in software testing on individual-level and system-wide level. AI adoption in testing can be done on individual level, where the QA specialists use AI as more of a personal assistant, and their goal is to, for example, increase their own productivity when performing their tasks. Another way of looking at AI adoption is the system-wide point of view, where an organization adopts AI on a larger scale, potentially changing the organization wide processes. When looking at quantitative measures in the studies for AI adoption in the section "The Current State of AI Adoption in Software Testing", the adoption rate was measured only on the individual level, usually among QA experts. There would be more potential for researching the adoption of AI in software testing also on the larger organizational level, and the cross-section of individual-level and system-wide adoption. According to Radhakrishnan et al \cite{Radhakrishnan2022UnderstandingApproach}, diffusion of innovations and technology-organization-environment framework (TOE) are the most prominent adoption models on a firm-level. Theory of planned behaviour (TBD), the technology acceptance model (TAM), or the unified theory of acceptance and use of tehnolgogy (UTAUT) could be potential theoretical frameworks to apply in the individual level \cite{Radhakrishnan2022UnderstandingApproach}. In literature, there is also another view of AI adoption: AI as a team member \cite{Hopf2024TheWork}. Hopf et al \cite{Hopf2024TheWork} have developed a "Transactive Intelligent Memory System (TIMS) as a new vision of collaboration between humans and IAs in hybrid teams".

Another potential research direction in AI in software testing is related to improved communication. For example, communication between software testers and developers is at times challenging, for example due to temporal \cite{Cruzes2016CommunicationTesting} or physical distance \cite{Taipale2007ObservingManagement}. Javed et al \cite{Javed2025ExploringExperiences} conducted a study in the context of learning and education, which confirmed that AI improves communication, task coordination, and project efficiency, making it a helpful tool for group work. .

Our findings show that there have been few peer-reviewed studies conducted within the real-world industry context related to AI in software testing. Of the 17 studies identified, only nine were peer-reviewed, while the remainder comprised theses and other literature.

The methodological gaps identified in the qualitative papers include a lack of research employing constructivist or interpretivist approaches, as well as a limited diversity in the qualitative methodologies utilized. It would be interesting to see more action research or ethnography studies. Instead of thematic analysis, other qualitative analysis methods, such as grounded theory could also be utilized. Also within thematic analysis itself, there are several different types of thematic analysis approaches. There could also be potential for interpretative phenomenological analysis, narrative analysis, and discourse analysis, that are rarely seen in software engineering research according to Lenberg et al \cite{Lenberg2024QualitativeGuidelines}. 

Adam Porter summarizes the need for additional research, and how it will impact AI adoption in the virtual round-table by Layman and Vetter \cite{Layman2024GenerativeTesting}:

\textit{"There will need to be a careful examination of our software testing needs and processes, a thorough identification of GenAI strengths and weaknesses, a widespread exploration  of specific use cases, and a data-driven comparison against existing solutions. We are only in the beginning stages of GenAI use. Much more experience and hard data will be needed before GenAI adoption becomes widespread."} 

In our view, this would apply to other AI technologies as well, not just generative AI.

\subsection{Limitations of the Study}

Our analysis employed a bottom-up approach, wherein we extracted actual use cases and benefits observed in real-life scenarios to achieve concreteness. We recommend additional research on the use cases and benefits on AI in software testing.

More actual benefits and use cases may exist in real world, than what we have reported in out study. We analyzed only the papers included in our sample. It should be also noted, that machine learning, deep learning, and other AI technologies were not used in search terms, and time range was limited to 2020+: instead of delving deep into the AI algorithms and technology, we were more interested in the more recent AI technologies, such as LLMs. 

It is also possible that some relevant papers were excluded based on titles or abstracts. According to  \cite{Brereton2007LessonsDomain}, it is difficult to evaluate the relevance of a study based on the abstract alone, so we can imagine, that evaluate the content of the paper is even more difficult based on the title only. 

Only two databases were utilized, but we selected ones that in our opinion provided a wide range of studies, Google Scholar and Scopus. In our case, grey literature, such as theses, were identified through Google Scholar, enriching our dataset for thematic analysis and providing valuable first-hand experiences from companies and specialists. However, it should be noted that Google Scholar should not be the only data source in literature review \cite{Haddaway2015TheSearching}. Scopus was selected due to it's selection of high quality papers from a wide cross section of different fields.

In order to keep this paper's scope manageable, we had to prioritize the themes that we wanted to report. Our next steps are to further develop these additional themes, and for example, provide a deeper analysis about the reasons behind low AI adoption rate.

\section{Conclusions}\label{sec5}

Our research questions in this study were:
\begin{itemize}
    \item RQ1: What kind of studies have been made in the industrial or business context regarding AI adoption in software testing?
    \item RQ2: How is AI utilized in software testing in the industry?
\end{itemize}

In our systematic mapping study to address RQ1, we found 17 recent studies since 2020 with industry context related to AI adoption in software testing. Nine of the studies were peer-reviewed, the rest either theses or other grey literature. Most of the studies were qualitative, which was partially influenced by our exclusion criteria. The number of studies overall is still quite small. There is a need for more empirical studies on AI in software testing in the industry setting, including, for example, field experiments, case studies, and action research.

With the thematic analysis of the 17 papers, we addressed our RQ2. We found that several use cases for AI in software testing have been documented in research. Secondly, we identified different scopes of AI adoption. AI is utilized at individual-level, where individuals use AI as an assistant to complete their tasks, and on system-wide level, where AI is used to automate large system-wide tasks. Thirdly, AI adoption in software testing still seems to be in very early stages, and the benefits drawn from the use cases can be vague or limited. 

We also confirmed that indeed expectations and reality do not yet meet in AI adoption. And in general, AI has not yet brought about a revolutionary transformation of software testing in the software industry. We compared the actual use cases and benefits with expected use cases and benefits, and connected the use cases to their respective actual and potential benefits. Time-savings as a benefit was observed in multiple studies, resulting from several use cases. Cost saving and increased job satisfaction were expected, but were not observed in the studies we analyzed. Overall, as documented by other studies, the expectations are high, but practical implementations are still quite far behind.

Additionally, we noticed a pattern in the data, where AI adoption in software testing was done with different scopes: as an individual-level or system-wide activity. In the case of individual-level adoption, AI is used by QA specialists in their own daily tasks, for example to boost their personal level productivity via code or document generation. In the case of system-wide adoption, AI is utilized in larger scale manner, such as generating test cases for the whole system, or as AI-based intelligent test automation.  

The sentiment that AI cannot replace testers, at least in the near future, was quite commonly shared by QA experts. But on the other hand, AI will impact the processes, workflows and roles, and will create new work, for example related to maintenance and monitoring of AI.

\bibliography{references}

\end{document}